\documentclass[12pt,a4paper]{article}
\pdfoutput=1
\usepackage{jheppub}

\usepackage{amssymb}
\usepackage{amsmath}
\usepackage{float}
\usepackage{graphicx}
\usepackage{amsfonts}
\usepackage{hyperref}
\usepackage{subfigure}
\newcommand{\be}{\begin{equation}}
\newcommand{\ee}{\end{equation}}
\newcommand{\bea}{\begin{eqnarray}}
\newcommand{\eea}{\end{eqnarray}}
\newcommand{\barr}{\begin{array}}
\newcommand{\earr}{\end{array}}

\def\beq{\begin{equation}}
\def\eeq{\end{equation}}
\def\be{\begin{equation}}
\def\ee{\end{equation}}
\def\bea{\begin{eqnarray}}
\def\eea{\end{eqnarray}}
\def\bg{\begin{eqnarray}}
\def\nd{\end{eqnarray}}

\title{Non-Canonical Inflation in Supergravity}

\author{Rhiannon Gwyn and Jean-Luc Lehners\\}
\affiliation{Max-Planck-Institute for Gravitational Physics (Albert-Einstein-Institute) \\ Am M\"{u}hlenberg 1, D-14476 Potsdam-Golm, Germany\\ 
{\tt rhiannon.gwyn; jean-luc.lehners@aei.mpg.de}}

\abstract{We investigate the effect of non-canonical kinetic terms on inflation in supergravity. We find that the biggest impact of such higher-derivative kinetic terms is due to the corrections to the potential that they induce via their effect on the auxiliary fields, which now have a cubic equation of motion. This is in contrast to the usual (non-supersymmetric) effective field theory expansion which assumes that mass-suppressed higher-derivative terms do not affect the lower-derivative terms already present. We demonstrate with several examples that such changes in the potential can significantly modify the inflationary dynamics. Our results have immediate implications for effective descriptions of inflation derived from string theory, where higher-derivative kinetic terms are generally present. In addition we elucidate the structure of the theory in the parameter range where there are three real solutions to the auxiliary field's equation of motion, studying the resulting three branches of the theory, and finding that one of them suffers from a singularity in the speed of propagation of fluctuations.}

\begin{document}

\maketitle
\section{Introduction}
The observational data available on the cosmic microwave background (CMB) \cite{Smoot:1992td, Komatsu:2010fb, Dunkley:2010ge, Story:2012wx}, including the latest PLANCK data \cite{Ade:2013zuv, Ade:2013uln}, baryon acoustic oscillations \cite{Percival:2009xn}, redshift \cite{Riess:1998cb, Perlmutter:1998np} and Hubble parameter measurements \cite{Riess:2011yx}, portrays a universe very close to being spatially flat, in which large-scale structure was generated from an almost scale-invariant power spectrum of primordial density perturbations with Gaussian distribution. A major contemporary challenge in cosmology is to find a convincing, self-consistent, explanation for this early state of the universe. The two most promising theories to date are a period of single-field plateau-type inflation \cite{Ade:2013uln}, and certain classes of two-field cyclic models of the universe \cite{Lehners:2013cka,Fertig:2013kwa}. Both theories fit the data well, subject to important assumptions: in the inflationary case, one assumes for instance that the right initial conditions for inflation were present for its onset, and that, after regulating the infinities that eternally inflating plateau models lead to, the naive predictions still hold (for contrasting views on these issues, see \cite{Ijjas:2013vea} and \cite{Guth:2013sya}). In cyclic models, the most important assumption is that a smooth, non-disruptive bounce from the contracting to the expanding phase is possible. In the present paper, we will only look at the inflationary case, and we will assume that the assumptions mentioned above are justified. Our concern will rather be with the (classical) dynamics during the inflationary phase.

In the absence of a UV-complete theory of quantum gravity, the dynamics of inflation is usually treated using an effective field theory (EFT) approach. Any given theory of inflation is then only valid up to some energy cut-off  $\Lambda$, with the physics above this scale being integrated out. The Lagrangian is organized into a series of kinetic and potential terms with increasing numbers of fields and/or derivatives, in such a way that successive terms are suppressed by increasing powers of the cut-off $\Lambda/M_{P}$ (expressed in Planck units). For non-derivative interaction terms, it is well known that, despite this suppression, the predictions of inflation are highly sensitive to certain higher-order terms in the series. For instance, the underlying UV completion can lead to large corrections to the inflaton mass via dimension 6 operators of the form $\frac{{\cal O}_6}{M_P^2} \sim \frac{{\cal O}_4}{M_P^2} \phi^2$, when ${\cal O}_4$ has a VEV of the order of the inflationary energy density. This is known as the $\eta$ problem and is a generic problem in supergravity theories of inflation, see e.g. \cite{Baumann:2009ni}. The underlying UV-complete theory can also lead to important corrections to the kinetic terms, resulting in higher-order corrections such as the square of the ordinary kinetic term, $(\partial \phi)^4$ \cite{Silverstein:2003hf, Franche:2009gk}. These higher-derivative kinetic terms can have dramatic effects and have been studied extensively, for instance in DBI inflation \cite{Silverstein:2003hf, Alishahiha:2004eh, Shandera:2006ax, Chen:2004gc, Chen:2005ad, Kecskemeti:2006cg}, in the context of the Horndeski action \cite{Horndeski:1974wa, RenauxPetel:2011sb, Ribeiro:2012ar} or within the EFT of inflationary perturbations \cite{Cheung:2007st, Senatore:2009gt, Creminelli:2010qf, Baumann:2011su, Achucarro:2012sm}. 

Whether the correct UV-complete underlying theory of inflation is string theory or some other high-energy theory is as yet unknown. Nevertheless, one may expect this theory to be supersymmetric for various reasons including unification of the electroweak and strong forces at high energies, the hierarchy problem \cite{Dimopoulos:1981zb, Dimopoulos:1981yj} and phenomenology \cite{Lukas:1998yy, Braun:2005nv}. The proper low-energy setting for a theory of inflation is then supergravity, and the goal of the present paper is to investigate a number of novel properties that the extension of the inflationary EFT framework to supergravity entails. An initial study of this extension was undertaken by Baumann and Green in \cite{Baumann:2011nm, Baumann:2011ws}, where they studied inflationary {\it perturbations} and focussed on possible non-Gaussian observational signatures. In that context, Baumann and Green argued that they could neglect the auxiliary fields. In the present paper, we will focus on the inflationary {\it background} -- in that situation, the auxiliary fields play a crucial role. In fact, it is the interplay of the higher-derivative kinetic terms and the auxiliary fields that leads to all of the new effects that we found, as we will briefly describe now.

The most important field in our study will be the auxiliary field $F$ of a chiral superfield $\Phi.$ Assuming that the auxiliary fields of the supergravity multiplet have already been eliminated via their equations of motion, the formula for the potential reduces to the expression
\be
V=-e^{K/3} K_{,AA^\star}FF^\star -e^{2K/3} [F (D_A W) +F^\star (D_A W)^\star] -3 e^KW W^\star. \label{FPotential}
\ee 
Here $K$ is the K\"{a}hler potential, $W(A)$ is the superpotential and $A=\Phi\mid_{\theta=\bar{\theta}=0}$ is a complex scalar representing the lowest component of the chiral superfield $\Phi$ (with $A$ containing the inflaton as one of its two real scalar field constituents); moreover, $D_A W = W_{,A} + K_{,A} W$ is the K\"{a}hler covariant derivative. 
Now substituting the standard solution of the auxiliary field equation 
\be
F=-e^{K/3}K^{,AA^\star} (D_A W)^\star \label{F_StandardSolution}
\ee
leads to the famous formula \cite{Cremmer:1978hn}
\be
V=e^K \left[ K^{,AA^\star} D_A W (D_A W)^\star -3 W W^\star \right]. \label{FamousPotential}
\ee 
Thus the potential is given as the difference between two manifestly positive terms (assuming the standard sign for the kinetic term, i.e. a positive K\"{a}hler metric $K_{,AA^\star}$), and hence these terms are larger (and in some cases significantly larger) in magnitude than the final potential. This simple observation turns out to have important consequences in our present context: we will add the leading higher-derivative correction term to the standard kinetic term for $A$ -- in supergravity this term is of the form \cite{Koehn:2012ar, Farakos:2012qu}
\be
T (\partial A)^2 (\partial A^\star)^2-2T e^{K/3} \partial A \cdot \partial A^\star FF^\star + T e^{2K/3} (FF^\star)^2 , \label{X2}
\ee
where $T$ has dimensions $[M^{-4}]$ and for now we take $T$ to be a small constant (when expressed in Planck units) \footnote{Here we are suppressing spacetime indices, so for instance $(\partial A)^2 \equiv \partial^\mu A \partial_\mu A$.}. As one can see from the above formula, adding a higher-derivative kinetic term in supergravity brings with it corrections to the lower-order kinetic terms (second term) as well as to the potential (third term). This is the first non-trivial consequence of supersymmetry: one cannot simply build up an effective field theory term by term, since each new term modifies some of the terms previously present! The second consequence comes from estimating the leading corrections by simply plugging in the solution for $F$ that is valid in the absence of the higher-derivative terms, namely Eq. \eqref{F_StandardSolution}, into Eq. \eqref{X2} (with $K_{,AA^\star=1}$):
\be
T (\partial A)^2 (\partial A^\star)^2-2T \partial A \cdot \partial A^\star e^{K} D_A W (D_A W)^\star + T [e^K D_A W (D_A W)^\star]^2 , \label{X2-2}
\ee
One would naively have expected that all corrections which the addition of the term \eqref{X2-2} leads to are of ${\cal O}(T)$ and that therefore, for small $T$ (say $T=1/100$), they can safely be ignored. However, as we can now see, the potential $V$ gets corrected not by a term $T V^2,$ but rather by a term proportional to $T[e^K D_A W (D_A W)^\star]^2.$ As argued below Eq. \eqref{FamousPotential}, this term can be significantly larger than $TV^2$ and thus the inclusion of higher-derivative {\it kinetic} terms can, and does, lead to important modifications of the {\it potential}. In the context of inflationary cosmology, this is evidently of importance, since the potential needs to be very flat over a significant field range. As we will see in more detail later on, such corrections have a crucial influence on the dynamics.  

In the present paper, we will develop the heuristic ideas above and present a quantitative analysis of the effects of the leading higher-derivative kinetic correction term in supergravity for several representative examples. In section \ref{structureoff} we will analyze the general structure of the theory, arguing analytically that the $T$-dependent correction to the potential will dominate over the correction to the kinetic term. We confirm this numerically by considering a number of explicit examples in section \ref{examples}, where we study the (inflationary or not) dynamics and their dependence on $T.$ The relationship with the formalism for finding inflationary attractor solutions is discussed in section \ref{attractor}. In section \ref{csblowup} we study the theory for negative $T$ where there can be three real solutions to the equation of motion for $F$. We find a blowup in the speed of sound $c_s^2$ for the branch which is smoothly connected to the $T > 0$ solution. We conclude with a discussion section. Appendix A contains the relevant stability analysis for the single-field solutions that we describe in the paper.


\section{Structure of the auxiliary field $F$}
\label{structureoff}
\subsection{Setup}

It was shown in \cite{Khoury:2010gb} how to construct ${\cal N} = 1$ supersymmetric actions involving higher-order kinetic terms of chiral superfields in flat superspace. This was generalized to ${\cal N} = 1$ supergravity, or curved superspace,  in \cite{Koehn:2012ar}, where the general Lagrangian for higher-derivative terms in supergravity was found to be\footnote{For closely related work, see also \cite{Buchbinder:1988yu, Buchbinder:1994iw, Banin:2006db, Brandt:1993vd, Brandt:1996au, Antoniadis:2007xc, Khoury:2011da, Farakos:2012qu, Koehn:2012te, Koehn:2013hk, Koehn:2013upa}}:
\begin{eqnarray}
\nonumber  \frac{1}{e} {\cal L} & = & \frac{1}{2} {\cal R}  - K_{A A^\star}  \partial A \cdot \partial A^\star + K_{A A^\star}  e^{K/3} F F^\star + e^{2K/3}\left  [ F(D_A W)  + F^\star (D_AW)^\star\right ] \\ \nonumber &&+ 3e^K W W^\star   + 16 (\partial A)^2 (\partial A^\star)^2 T_{A A^\star A A ^\star}- 32 e^{K/3} F F^\star (\partial A \cdot \partial A^\star) T_{A A^\star A A^\star} \\  \label{genaction} &&+ 16 e^{2K/3} F^2 (F^\star)^2 T_{A A^\star A A^\star},
\end{eqnarray}
where $ T_{A A^\star A A^\star} $ is the lowest component of the tensor superfield $T_{i j k\star l\star}$  \cite{Koehn:2012ar} which controls the higher-derivative kinetic terms. It is a function of $A$ and can contain spacetime derivatives of $A$ as long as all spacetime indices are contracted. In this paper we will choose $$T_{AA^\star A A^\star}= T (K_{,AA^\star})^2,$$ with $T$ being a constant. Given that the action is quartic in the auxiliary field $F$, one sees immediately that this implies a cubic equation of motion for $F.$ This can lead, in certain regions of the parameter space, to there being three real solutions for $F$, which we explore in the following. An initial study of the action (\ref{genaction}) was undertaken in \cite{Koehn:2012ar}, in which the solutions of the cubic for $F$ and the behavior of a general system in the limiting cases where $T \rightarrow 0$ and $T \rightarrow \infty $ were considered. Here we expand this analysis significantly, considering the details of the solutions for $F$ for all values of $T$, examining the relative importance of the kinetic and potential contributions, and exploring details of the theory numerically for several explicit examples. 

We start by writing out the equation of motion for $F$:
\begin{eqnarray}
\label{cubic1}
0 & = & K_{A A^\star} F + e^{K/3} (D_A W)^\star + 32 F T_{A A^\star A A^\star} \left (e^{K/3} F F^{ \star} - \partial A \cdot \partial A^\star \right ) .
\end{eqnarray}
Multiplying (\ref{cubic1}) by $F^\star$ we see that $(D_AW)^\star F^\star$ must be real, which implies 
\begin{eqnarray}
F^\star & = & \frac{D_A W}{(D_A W)^\star} F. \label{Freal}
\end{eqnarray}
Then the Lagrangian and cubic become
\begin{eqnarray}
\nonumber \frac{1}{e} {\cal L} & = & \frac{1}{2} {\cal R} - K_{A A^\star} (\partial A \cdot \partial A^*) + K_{A A^\star} e^{K/3} \frac{D_A W}{(D_A W)^\star}F^2  + 2 e^{2K/3} F(D_A W)  \\ \nonumber  &&+ 3 e^KW W^\star   + 16 (\partial A)^2(\partial A^\star)^2 T_{A A^\star A A^\star} + 16 e^{2K/3} F^4 \left ( \frac{D_A W}{(D_A W)^\star}\right ) ^2T_{A A^\star A A^\star} 
 \\ \label{Aaction} &&
  - 32 e^{K/3} F^2 \frac{D_A W}{(D_A W)^\star} (\partial A \cdot \partial A^\star) T_{A A^\star A A^\star} \\
\nonumber 0 & = & K_{A A^\star} F + e^{K/3} (D_A W)^\star  + 32 \left [ e^{K/3} \frac{D_A W}{(D_A W)^\star} F^3 - \partial A \cdot \partial A^\star  F \right ] T_{A A^\star A A^\star}.\\
\end{eqnarray}
In order to proceed, we must solve for the auxiliary field $F$ and plug it back into the Lagrangian above. The solutions are qualitatively different depending on whether $T$ is positive or negative, and hence we will deal with these two cases separately (the negative $T$ case will be dealt with in section \ref{csblowup}).

\subsection{Solving the cubic: $T\geq0$}
\label{Tpos}
The cubic can be rewritten in standard form as \cite{Koehn:2012ar}
\begin{eqnarray}
0 & = & F^3 + p F + q\\
p & = & e^{- K/3} K^{AA^\star}\frac{(D_A W)^\star}{D_A W}  \left [\frac{1}{32 T}-  K_{AA^\star} \partial A \cdot \partial A^\star  \right ]\\
q & = & \frac{1}{32 T}\frac{(K^{AA^\star})^2(D_A W)^{\star 2}}{D_A W} 
\end{eqnarray}
We will often specialize to the case where we consider only one real scalar of $A,$ e.g. we write $A = \frac{1}{\sqrt{2}}\phi$ with $\phi$ being real and denoting the inflaton. In the cosmological context, we will treat $\phi$ as being a function of time only and use the standard abbreviation $$X \equiv \frac{1}{2}\dot{\phi}^2.$$ In this case, with the superpotential being a so-called real holomorphic function of $A$ (i.e. a function of $A$ where the series expansion contains only real coefficients), Eq. \eqref{Freal} implies that $F$ must be real. For $T \geq 0$ and $p, q \in {\mathbb R}$ there is generally only one real solution to the cubic for $F$, given by \cite{Koehn:2012ar}
\begin{eqnarray}
F & = & (- \frac{q}{2}+ \sqrt{D})^{1/3} + (- \frac{q}{2} - \sqrt{D})^{1/3},
\end{eqnarray}
where $D = \frac{q^2}{4} + \frac{p^3}{27}$ is the discriminant. This can also be compared to the $T = 0$ solution for $F$, which is just
\begin{eqnarray}
\label{T0F}
F_{T\,=\,0} & = & - e^{K/3} K^{A A^\star} (D_A W)^\star .
\end{eqnarray}
Thus, for these cases, the consistency condition (\ref{Freal}) reduces the number of permitted branches of the theory to only one. (In section \ref{csblowup} we will study examples where more branches are allowed.)
  
\begin{figure}[htp]
\includegraphics[width=0.5 \textwidth]{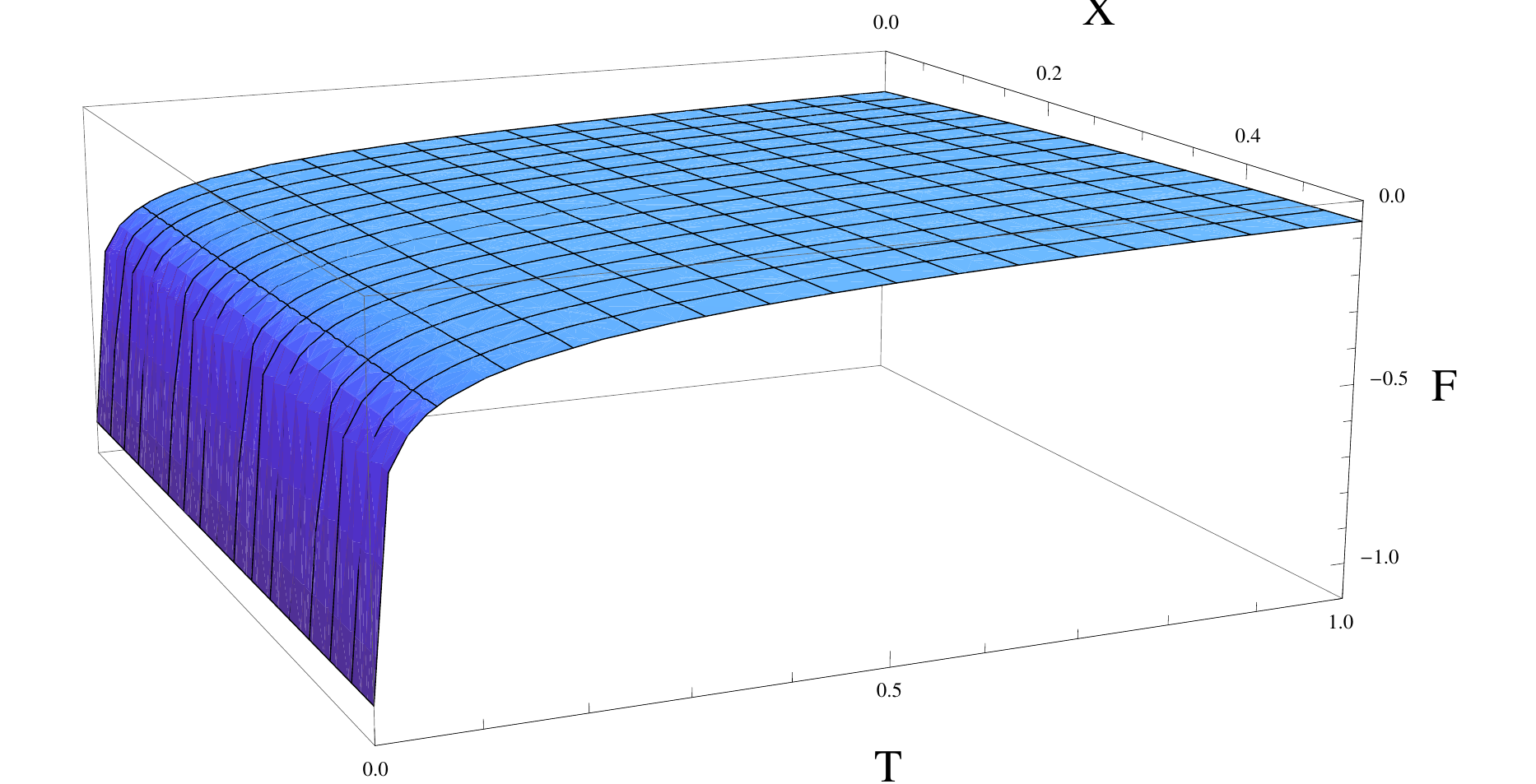}
\includegraphics[width=0.5 \textwidth]{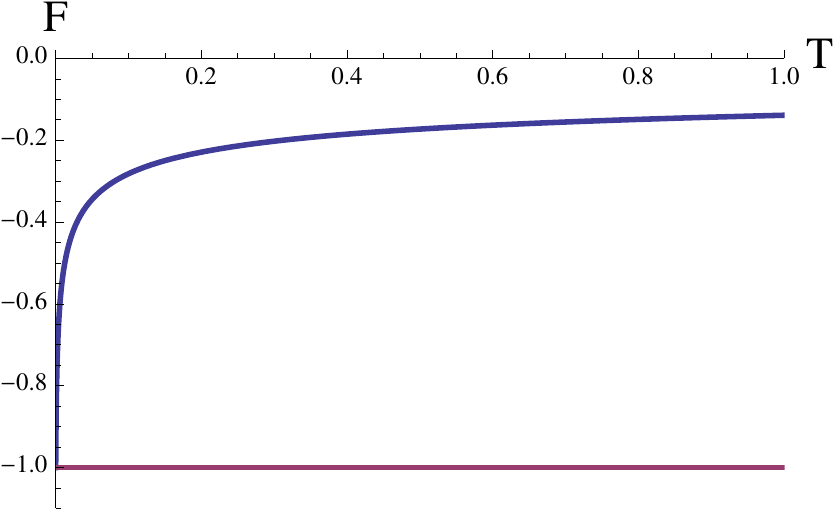}
\caption{The auxiliary field $F$ as a function of $X$ and $T$, for $T>0$ and with $C_1 = C_2 = 0.1$. We see that $F$ is a negative function, which, starting from its standard $T=0$ solution (here arbitrarily set at the value $F=-1$), approaches zero asymptotically as $X$ and $T$ increase. 
}
\label{FposT}
\end{figure}

To study the dependence of $F$ on $X$ and $T,$ note that the cubic is of the form (momentarily setting $K_{AA^\star}=1$)
\bg
F^3 + \frac{C_1(\phi) F}{32 T} (1 + 32 TX) + \frac{C_2(\phi)}{32 T}=0,
\label{cubic_general}
\nd
so that we can write
\bg
p& = &  \frac{C_1(\phi)}{96 T} (1 + 32 XT),\\
q & = & \frac{C_2(\phi)}{32T},
\nd
where $C_{1,2}$ are functions of $\phi$ only. Thus, for given values of $C_{1,2}$, the $X$ and $T$ dependence of $F$ will be the same in each case; see Figure \ref{FposT}. As is evident from the figure, $F$ starts out at $T=0$ at its standard value \eqref{T0F} and then approaches zero as $T$ and $X$ increase. There is one immediate consequence of this fall-off towards zero, namely that the potential becomes more and more negative with increasing $T$ and $X$ (as should be evident from Eqs. \eqref{FPotential} and \eqref{FamousPotential}). An explicit example of this general trend is provided by the supergravity embedding of DBI inflation, as described in \cite{Koehn:2012np}.

We will make this more precise now by calculating the perturbative corrections induced when $T$ is non-zero. By considering the action (\ref{Aaction}), we can see that the higher-derivative terms, parametrized by $T$, enter the action in two distinct ways: via the higher-power kinetic term $16 T X^2$, and via the $T$-dependence of $F$, which leads to corrections to the kinetic terms as well as significant corrections to the potential. In \cite{Koehn:2012ar} a small $T$ expansion of $F$ was given, using the approximation $q \ll p^{3/2}$. In this limit,
\begin{eqnarray}
\label{smallTapprox}
F_{q \ll p^{3/2}} & \approx  &- \frac{q}{p} + \frac{q^3}{p^4} - \frac{3 q^5}{p^7} +  ...\\
\nonumber   & \approx &  - e^{K/3} K^{A A^\star}  (D_A W)^\star \\ \nonumber && + 32 T e^{K/3} K^{A A^\star} (D_A W)^\star \left ( K^{A A^\star} e^K |D_A W|^2 - K_{AA^\star}\partial A \cdot \partial A^\star\right) \\ \nonumber && 
- 32^2 T^2 e^{K/3} K^{AA^\star} (D_A W)^\star  \left (3 e^K K^{A A^\star} |D_AW|^2  - \partial A \cdot \partial A ^\star K_{A A^\star} \right ) \times \\ \nonumber && \left (e^K K^{A A^\star} |D_AW|^2  - \partial A \cdot \partial A^\star K_{A A^\star} \right ) .
\end{eqnarray}
This gives $T$-dependent corrections up to ${\cal O}(T^2)$ to the standard result for $T = 0$ (\ref{T0F}).
In order to determine when such corrections are significant, we must determine the conditions for $q / p^{3/2} $ to be large:
\begin{eqnarray}
\frac{q^2}{p^3} & = & \frac{32 T e^K K^{A A^\star} D_A W (D_A W)^\star}{ (1 - 32 T K_{AA^\star} \partial A \cdot \partial A^\star)^3}.
\end{eqnarray}
This expression can be large even when $T$ is small, as long as $32 e^K K^{A A^\star} |D_A W|^2$ is sufficiently big. We would like to point out that this can be the case even when the original potential $$V_{T=0} = e^K (K^{A A^\star} |D_A W|^2 - 3|W|^2)$$ is small, since the potential is given by the difference of two potentially large terms. Furthermore, the resulting corrections to the action will affect the kinetic and potential term contributions in different ways. Writing the Lagrangian as
\begin{eqnarray}
\frac{1}{e}{\cal L} &=& \frac{1}{2}{\cal R} - K_{A A^\star} \partial A \cdot \partial A^\star [ 1 + \Delta K  ] +16 T (\partial A)^2 (\partial A^\star)^2 \nonumber \\ && - e^K (K^{A A^\star} |D_AW|^2 - 3 |W|^2) [ 1+ \Delta V ],
\end{eqnarray}
to quadratic order in $T$, we obtain
\begin{eqnarray}
\Delta V & = &  -\frac{16 T}{V} e^K (K^{A A^\star})^2 |D_A W|^4  + \frac{32^2 T^2}{V} e^{2 K} (K^{A A^\star})^3 |D_A W|^6\\
\Delta K & = & 16 T (- K_{A A^\star}\partial A \cdot \partial A^\star + 2 e^K K^{AA^\star} |D_A W|^2 ) \\ \nonumber && - 64 T^2 e^K  K^{A A^\star} |D_A W|^2 (- 16 K_{AA^\star} \partial A \cdot \partial A^\star + e^K K^{A A^\star} |D_A W |^2).
\end{eqnarray}
The case of greatest interest to us is where $e^K K^{A A^\star} |D_A W|^2$ is significantly larger than $V$ - then, for small field velocities we have 
\begin{eqnarray}
\left | \frac{\Delta K}{\Delta V}\right | & \approx & \frac{2V}{e^K K^{A A^\star} |D_A W|^2}.
\end{eqnarray}
Thus, when the higher-derivative term modifies the lower-derivative terms significantly, it tends to affect the potential much more than the (ordinary) kinetic term. We will explore these effects for specific examples in detail in section \ref{examples}, but one can immediately see the effect on various potentials (which are defined below), by looking at Figs. \ref{FigurePot1} and \ref{FigurePot2}. As is already evident by eye, even for small $T$ the potentials are dramatically altered in certain field ranges.


\section{Examples}
\label{examples}
We will now consider several explicit examples in order to illustrate the general effects described above. Specifically, we will look at the representative cases listed in table \ref{examplestable} below. We are using the notation that $X=\frac{1}{2}\dot\phi^2, \, Y = \frac{1}{2}\dot \xi^2$ and we are taking $c$ to be a positive real number. Note that in the flat potential case $P_{Fl}$, the superpotential $W_{Fl}$ is chosen such that the uncorrected ($T=0$) potential is equal to a constant $V_0$ along the $\xi=0$ line \cite{Lehners:2002tw}.


\begin{table}
\caption{Details for four explicit examples.}
\label{examplestable}
\begin{tabular}{c || c}
 {\bf Canonical K\"ahler potential} &  {\bf Quadratic K\"ahler potential}\\
\hline \\
$
\begin{array}{rcl}
K_{CanK} & = & A^\star A \\
W_{CanK} & = & e^{c A} \\
A_{CanK} & = & \frac{1}{\sqrt{2}} (\phi + \imath \xi)\\
K_{A A^\star}&  = & 1\\
e^K & = & e^{\frac{(\phi^2 + \xi^2)}{2}}\\
\partial A \cdot \partial A^\star & = &  - (X + Y)\\
D_A W_{CanK} & = & \left ( c + \frac{1}{\sqrt{2}} ( \phi - \imath \xi) \right ) e^{\frac{c (\phi + \imath \xi)}{\sqrt{2}}}\\
 \end{array}
 $
 & 
 $\begin{array}{rcl}
 K_{QuadK} & = & - \frac{1}{2} (A - A^\star)^2\\
W_{QuadK} & = & e^{c A}\\
A_{QuadK} & = & \frac{1}{\sqrt{2}} (\phi + \imath \xi)\\
K_{A A^\star}&  = & 1\\
e^K & = & e^{\xi^2}\\
\partial A \cdot \partial A^\star & = &  - (X + Y)\\
D_A W_{QuadK} & = &  ( c - \sqrt{2} \imath \xi ) e^{\frac{c (\phi + \imath \xi)}{\sqrt{2}}}
 \end{array}$
\end{tabular}

\vspace{1 cm}

\begin{tabular}{c || c}
 {\bf Flat potential} &  {\bf Linear superpotential}\\
\hline \\
$
\begin{array}{rcl}
K_{Fl} & = &- 4 \ln (A + A^\star)\\
W_{Fl} & = & 2 \sqrt{\frac{V_0}{3}} A^2\left (A^{\sqrt{3}} - A ^{- \sqrt{3}} \right )  \\
A_{Fl} & = & e^{\frac{1}{\sqrt{2}} \phi} + \imath \frac{\xi}{\sqrt{2}}\\
K_{A A^\star}&  = &  e^{- \sqrt{2}\phi} \\
e^K & = & (2 e^{\frac{\phi}{\sqrt{2}}})^{-4}\\
\partial A \cdot \partial A^\star & = &  - (e^{\sqrt{2}\phi}X  + Y)\\
D_A W_{Fl} & = &4 \sqrt{\frac{V_0}{3}} A Q_- (1 - e^{- \phi/\sqrt{2}} A) \hspace{.17cm}\\&&+ 2 \sqrt{V_0} A Q_+\\
Q_\pm & = & A^{\sqrt{3}} \pm A^{- \sqrt{3}}
 \end{array}
 $
 & 
 $\begin{array}{rcl}
 K_{Lin} & = & - \frac{1}{2} (A - A^\star)^2\\
W_{Lin} & = & c \sqrt{2} A\\
A_{Lin} & = & \frac{1}{\sqrt{2}} (\phi + \imath \xi)\\
K_{A A^\star}&  = & 1\\
e^K & = & e^{\xi^2}\\
\partial A \cdot \partial A^\star & = &  - (X + Y)\\
D_A W_{Lin} & = &  c \sqrt{2} (1 - \imath \xi \phi + \xi^2)
 \end{array}$
\end{tabular}\\
\end{table}

For the four examples, the corresponding Lagrangians are easily obtained by plugging the potentials listed in the table into Eq. \eqref{genaction}. In our numerical studies, we will focus on the case where only the inflaton field $\phi$ is dynamical, and the second scalar $\xi$ remains fixed at $\xi=0.$ In that case, the auxiliary field takes only real values, and the respective matter Lagrangians simplify to the following expressions:
\begin{eqnarray}
\label{PCanK} P_{CanK} (X, \phi) & = & X + 16 X^2 T + 32 e^{\phi^2/6} F^2 X T + 16 e^{\phi^2/3} F^4 T\\ \nonumber &&  + e^{\phi^2/6} F^2 + 2 e^{\phi^2/3} F \left ( c + \frac{\phi}{\sqrt{2}} \right ) e^{\frac{c \phi}{\sqrt{2}}} + 3 e^{\phi^2/2} e^{\sqrt{2} c \phi}, \label{LagCanK} \\
P_{QuadK} (X, \phi) & = & X + 16 X^2T + 32F^2 X T + F^2 + 16 F^4 T + 2 c F e^{\frac{c \phi}{\sqrt{2}}} \label{PQuadK}\\ \nonumber && + 3 e^{\sqrt{2} c \phi}, \\
P_{Fl} (X, \phi) & = & X  + 16 X^2 T + 32 |F|^2 X T (2 e^{\frac{\phi}{\sqrt{2}}})^{-4/3} (e^{\phi/\sqrt{2}})^{-2} \label{PFl} \\ \nonumber && + 4 |F|^2 (2 e^{\phi/\sqrt{2}})^{-10/3} + \frac{V_0}{4} Q_-(\phi)^2 + 16 (|F|^2)^2 T (2 e^{\phi/\sqrt{2}})^{-8/3}(e^{\phi/\sqrt{2}})^{-4} \\ \nonumber && + 2 F (2 e^{\phi/\sqrt{2}})^{-5/3} \sqrt{V_0} Q_+ (\phi) , \\
\label{PLin}P_{Lin}(X, \phi) & = & X +  F^2 + 3  c^2 \phi^2 + 2 \sqrt{2} c F + 16 X ^2 T + 32 F^2 X T \\ \nonumber && + 16 F^2 F^{\star 2} T. \label{LagLin}
\end{eqnarray}
The stability of the potentials around the $\xi=0$ line is studied in Appendix A.


\subsection{Attractor Solution and Numerical Trajectories}
\label{trajectories}
\label{attractor}

A general formalism for inflation in the case of Lagrangians of the form $P(X, \phi)$ was given in \cite{Franche:2009gk}. It was shown there that when inflation takes place (i.e. when a set of generalized slow-roll coordinates is much smaller than one), it is an attractor in $(\phi, \Pi)$ phase space. Since these results apply in our case as well, we will briefly review the conditions for such a non-canonical inflationary attractor to exist.

The scalar field equation of motion arising from a Lagrangian ${\cal L} = a^3(t) P(X,\phi)$ can be written in the compact form 
\begin{eqnarray}
\dot \Pi = - 3 H \Pi + P_\phi,
\end{eqnarray}
where $\Pi = \dot \phi P_X = \sqrt{2X} P_X$ is the canonical momentum.  We can rewrite this in terms of the generalized slow-roll parameters \cite{Franche:2009gk}
\begin{eqnarray*}
\epsilon & = & \frac{3X}{\rho} P_X;\\
\eta_X & = & \epsilon - \frac{\dot X}{2HX};\\
\eta_\Pi & = & \epsilon - \frac{\dot \Pi}{H \Pi},
\end{eqnarray*}
finding
\begin{eqnarray}
\Pi & = & \frac{P_\phi}{3H} \left (1 + \frac{\epsilon + \eta_\Pi}{3} \right ) ^{-1}\\
\label{atteqn} \Rightarrow \Pi_{\mathrm{inf}} & = & \frac{P_\phi}{3 H}. \label{infsolution}
\end{eqnarray}
Note that there is no approximation in the expression for $\Pi$. However, in the case that $\epsilon$ and $\eta_\Pi$ are both small, or at least $\epsilon - \eta_\Pi \ll 1$, we obtain the inflationary solution  $\Pi \approx \Pi_{\mathrm{inf}}.$ To see that it is an attractor precisely when the inflationary parameters are small, we follow the analysis of \cite{Franche:2009gk}: for $$ \Pi = \Pi_{inf} ( 1 + \delta \Pi),$$ and autonomous equations
\begin{eqnarray}
\phi' & = & \frac{\Pi}{P_X H};\\
\Pi' &=&  - 3 \left  [\Pi - \frac{P_\phi}{3H} \right ],
\end{eqnarray}
we have
\begin{eqnarray}
\frac{\delta \Pi'}{\delta \Pi} & = & - 3 \left [1 + \frac{\epsilon - \eta_\Pi}{3} - \frac{c_s^2 \sqrt{2X}}{3H} \frac{P_{X \phi}}{P_X} + \frac{P_\phi \sqrt{2X}}{6 \rho H} \right ] \\
 & = & - 3 \left [1 + \frac{1}{3} (\eta_\Pi - \epsilon) c_s^2 + \frac{1}{3} (\epsilon - \eta_X) + \frac{\epsilon}{3} + \frac{\epsilon - \eta_\Pi}{3} \right ]\\
 & = & - 3 + {\cal O} (\epsilon, \eta_X, \eta_\Pi).
\end{eqnarray}
Note that the last two terms in the first line have the opposite sign compared to the corresponding terms in \cite{Franche:2009gk}; this is because of the sign chosen for $\dot \phi = \pm \sqrt{2X}$, where here we have chosen the plus sign.  From the final expression, we see that for small inflationary parameters, the perturbations $\delta \Pi$ from $\Pi_{inf}$ decay as $\delta \Pi \sim e^{-3 N}$. This is a localized attractor, in the sense that it is an attractor only in the region of phase space where the inflationary parameters are small.

\begin{figure}[h!]
\vskip -3mm
\centering
\includegraphics[width= 0.49\linewidth]{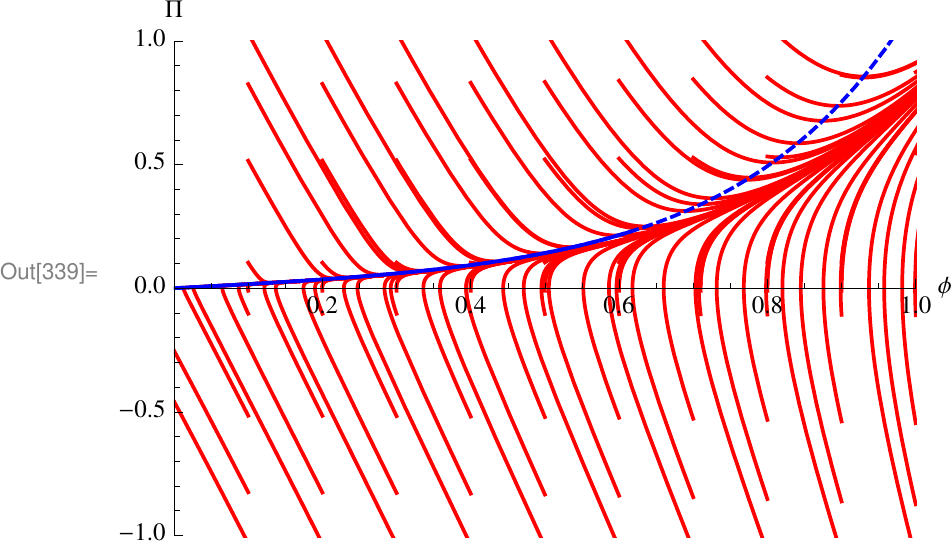}
\includegraphics[width= 0.49\linewidth]{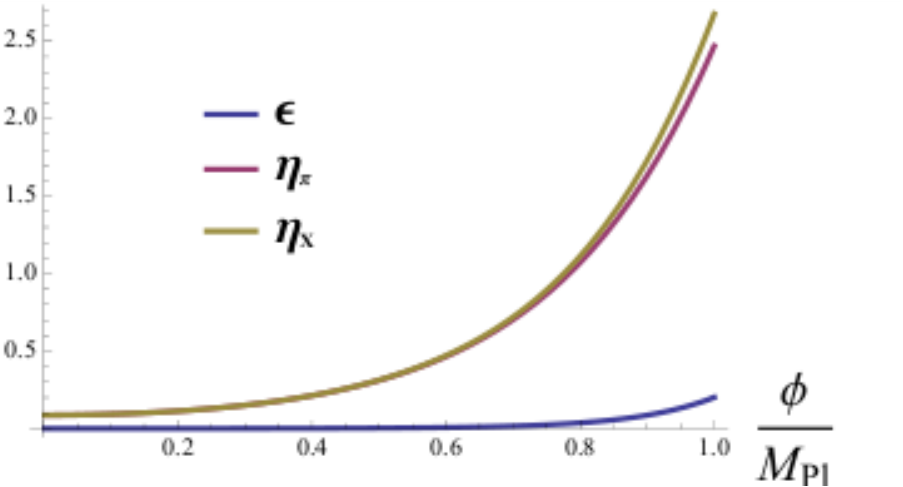}
\caption{The blue line indicates the inflationary attractor  (\ref{infsolution}) for  $P_{Fl}(X, \phi)$ with $V_0=10, \, T=10^{-3}$ (left). The line is solid until $\eta_X$ and $\eta_\Pi$ become greater than $0.5,$ and dashed from then on. The figure on the right shows the behavior of the generalized slow-roll parameters, evaluated along the blue solid/dashed line  (\ref{infsolution}).}
\label{FwrtX}
\end{figure}

The attractor behavior of the inflationary solution (\ref{infsolution}) is demonstrated below for $P_{Fl}(X, \phi)$ in the case where $V_0 = 10, T = 10^{-3}$, see Fig. \ref{FwrtX}. The red lines are a set of trajectories spanning a range of initial conditions (the initial conditions can be inferred from the starting points of the red lines). The blue line is the solution to  (\ref{infsolution}); it is solid up to $\phi \approx 0.612$ where $\eta_X$ and $\eta_\Pi$ become greater than $0.5,$ and dashed otherwise. As is evident from the figure, all trajectories that start in the vicinity of the attractor approach it quickly and then follow it until the slow-roll conditions are no longer satisfied. The behavior of the generalized slow-roll parameters, evaluated along (\ref{infsolution}), is shown on the right. For this specific example, we find that inflation lasts for a number of e-folds $N_e \approx 40 - 100$ for trajectories beginning at small $\phi$.

The Lagrangians given above in Eqs. \eqref{LagCanK} - \eqref{LagLin}, while very complicated to analyze analytically, can be studied numerically. For an FRW background, with $ds^2 = - dt^2 + a(t)^2 d {\bf x}^2$ and the field $\phi = \phi(t)$ homogeneous in space, we have to solve the coupled system of equations
\bg
0 & = & \ddot \phi(t)(P_X + 2X P_{XX}) + 2X P_{X \phi} + 3 H(t) \dot \phi(t) P_X - P_\phi, \\
0& = & \dot H(t) + X P_X ,
\nd
where we use the standard notation that a subscript indicates a derivative w.r.t. the corresponding quantity. We choose $c$ or $V_0$ and a fixed value of $T$ in each case with the hope of finding a noncanonical inflationary solution. Note that because of the $X$ dependence in $F$, the coefficient of the canonical kinetic term $X$ in the action is not in general canonically normalized.\footnote{Although it is possible to transform to the canonically normalized variable numerically, it cannot generally be done analytically, and as it does not affect the overall trends being studied and compared in the examples we consider here, we leave the normalization as is.}

\begin{figure}[H]
\centering
\mbox{\subfigure{\includegraphics[width=0.49 \linewidth]{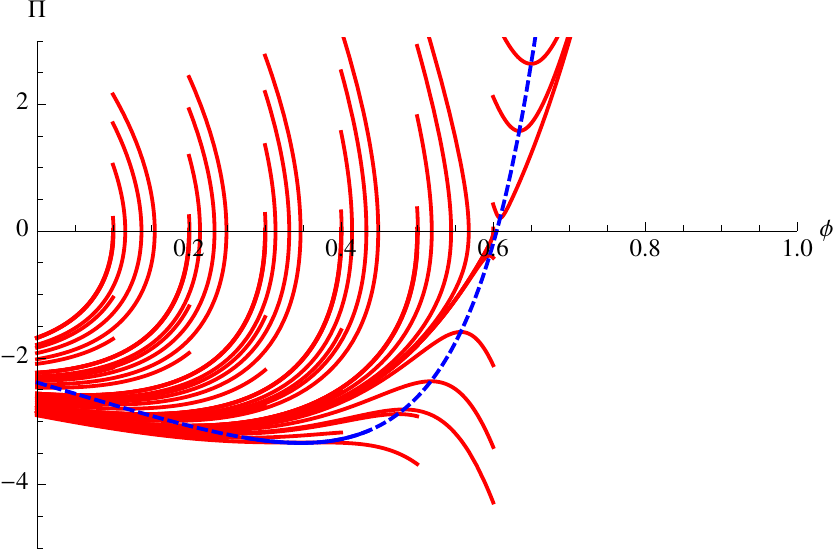}}\quad
\subfigure{\includegraphics[width=0.49 \linewidth]{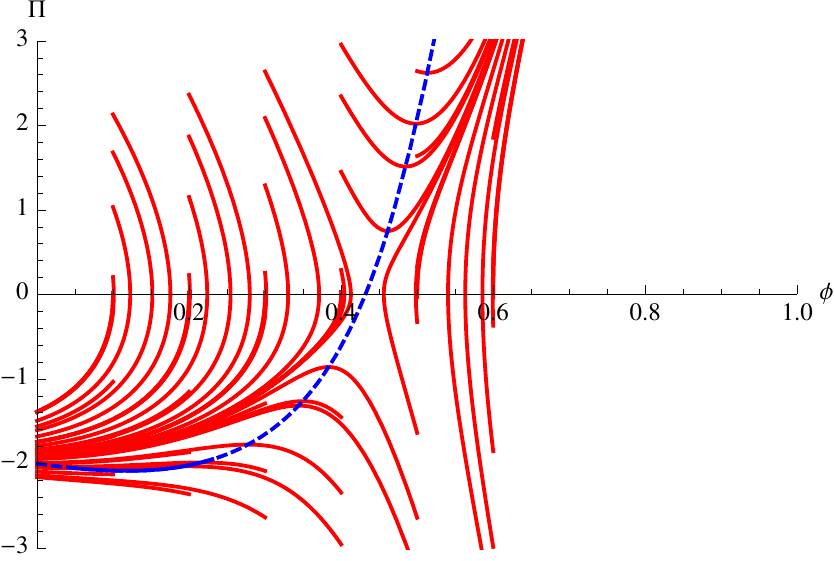}}}
\caption{Trajectories in $(\Pi, \phi)$ phase space for $c=3$ and $T = 0.01$ with $P_{CanK}(X, \phi)$ (left) and $P_{QuadK}(X, \phi)$  (right).  The blue line is the slow-roll approximation, which is a solid line until one of the generalized slow-roll trajectories reaches $1$.}
\label{PiTrajCanKPot}
\end{figure}

\begin{figure}[H]
\centering
\includegraphics[width=0.49 \linewidth]{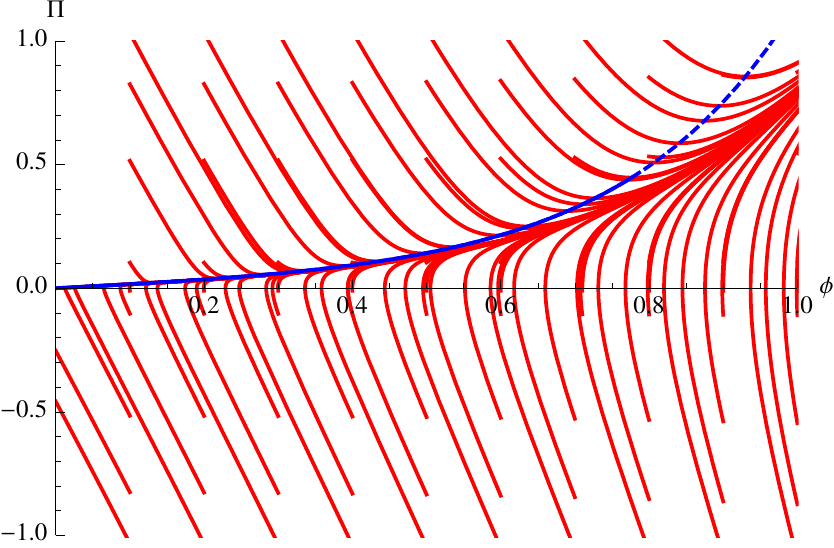}
\caption{Trajectories in $(\Pi, \phi)$ phase space for $P_{Fl}(X, \phi)$ with $V_0=10$ and $T = 0.0001$.} 
\label{PiTrajFlatPot}
\end{figure}

The plots for three of the examples considered in the previous section are given in Figs. \ref{PiTrajCanKPot} and \ref{PiTrajFlatPot} for the Lagrangians $P_{CanK}$, $P_{QuadK}$ and $P_{Fl}$ respectively. In each case, we plot the trajectories in the field amplitude/field momentum plane for a range of initial conditions that can be read off from the starting points of the red lines on the graphs.  Also indicated on the graphs is the slow-roll attractor solution Eq. (\ref{atteqn})  - it is drawn as a solid blue line when the slow-roll conditions are satisfied and a dotted line otherwise. As expected from the analysis above, this is an attractor when the slow-roll parameters are small, and the focussing of the trajectories around the solid blue line confirms this expectation. For the parameters shown, the potentials are stable around $\xi = 0$ until $\phi \sim 0.6$ for $P_{CanK}$ and $P_{QuadK}$, while $P_{FL}$ is generally unstable around $\xi = 0$ for small $\phi$ - see Appendix A.

We should point out that the linear superpotential leads to a very steep and generally negative potential, and is not particularly interesting for the study of inflationary trajectories. This example however illustrates one important facet of the present setting: if a potential is already unstable, the higher-derivative $T$-dependent corrections generally tend to make it even more unstable. This feature may be interesting in ekpyrotic models, where a steep negative potential is required \cite{Khoury:2001wf, Lehners:2008vx}. However, in the inflationary context, the higher-derivative terms play a more welcome role when the potential curves upwards too steeply (i.e. when $V_{,\phi\phi}>0$). As we will discuss next, in those cases the higher-derivative terms have two benefits: the potential is usually lowered, and the higher-derivative kinetic term tends to slow the field down.

\subsection{Effect of the noncanonical kinetic terms on the dynamics and on the potential}
\label{kinetictermseffect}
\begin{figure}[h!]
\vskip -3mm
\centering
\mbox{\subfigure{\includegraphics[width=0.49 \linewidth]{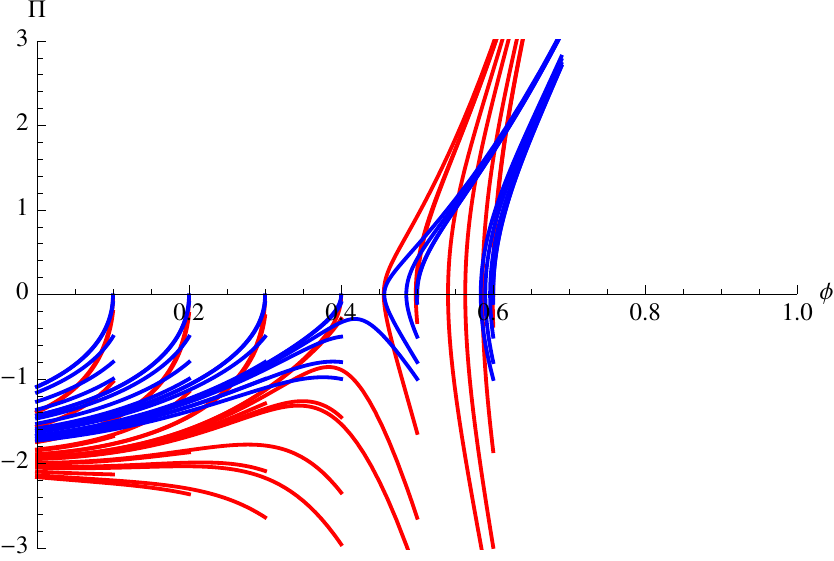}}\quad
\subfigure{\includegraphics[width=0.49 \linewidth]{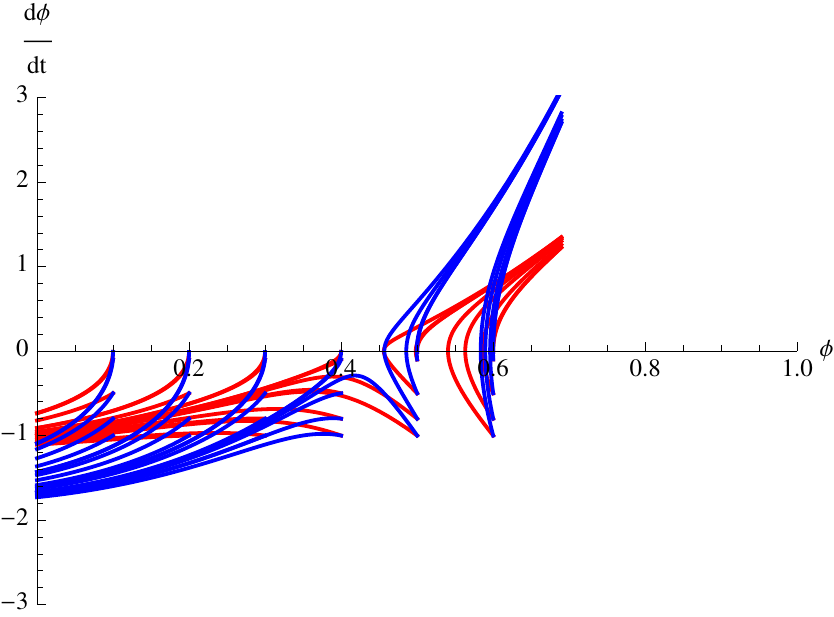}}}
\caption{Trajectories in $(\Pi, \phi)$ (left) and $(\dot \phi, \phi)$ (right) phase space for $P_{QuadK}(X, \phi)$ with $c = 3$ and $T = 0.01$. The same trajectories for $P_{QuadK}(X, \phi)$ with non-canonical kinetic terms switched off are in blue. } \label{FigNCvsC2}
\end{figure}

\begin{figure}[H]
\centering
\includegraphics[width=6cm]{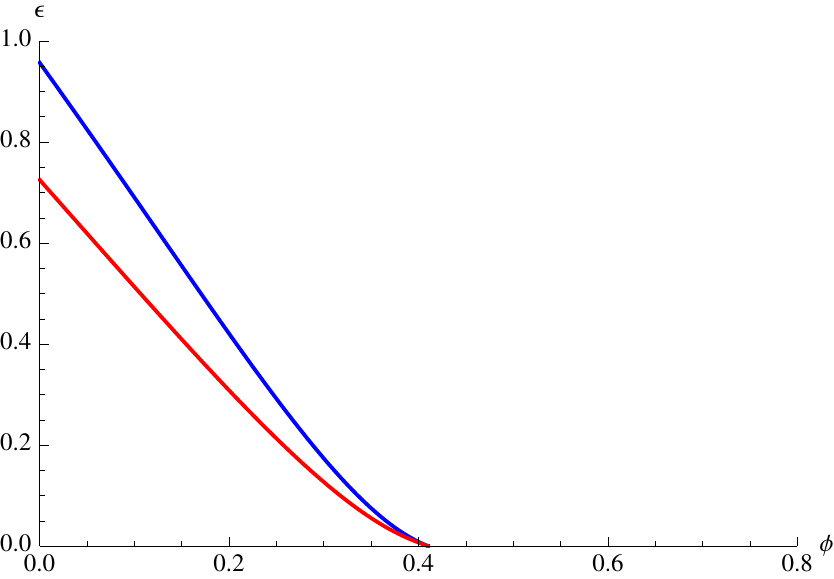}
\caption{$\epsilon(\phi)$ for $P_{QuadK}(X, \phi)$ with $c = 3$ and $T = 0.01$ for a single trajectory. The same trajectories for $P_{QuadK}(X, \phi)$ with non-canonical kinetic terms switched off are in blue. }\label{FigEps2}
\end{figure}

To study the effect of the noncanonical kinetic terms on the dynamics, we can compare the trajectories above to those for the action with {\it canonical} kinetic term, but where we keep the $T$-dependent potential. In other words, we compare $P_{NC}(X, \phi)$ with $P_C(X, \phi) = X + P_{NC}(0, \phi)$. Comparing the trajectories that we obtain in each case will tell us what the effect of the noncanonical kinetic terms is.  Looking at the resulting trajectories, we see that the canonical trajectories in $(X, \phi)$ are steeper, while the trajectories in $(\Pi, \phi)$ are less steep because of the effect of $P_X$. This is clearly seen in the representative plots for the Quadratic K\"ahler potential, Figure \ref{FigNCvsC2}. Thus, as expected from the intuition gained by studies of DBI inflation, we can see that the higher-derivative kinetic term slows the inflaton down. Correspondingly, in each case the slow-roll parameters are reduced, as shown in Figure \ref{FigEps2} (the other cases are very similar).

All these plots demonstrate that the higher-derivative kinetic term has a noticeable effect on the dynamics. However, the most dramatic effect is on the shape of the potential itself, as already discussed in section \ref{Tpos}. For our examples, given the solutions for the auxiliary field $F$, we can plot the respective potentials by setting $X$ to zero, i.e. we are interested in $V=-P(X=0, \phi)$. In each case, this is found to be highly sensitive to increases in $T$, with the potential rapidly flattening or changing in gradient until it becomes negative and steep. Obtaining a flat or almost flat potential is thus a matter of fine-tuning $T$, as is evident from the plots below in Figs. \ref{FigurePot1} and \ref{FigurePot2}. Moreover, it is generally not possible to tune the potential to be flat over a large field range, which, given past experience with the search for inflationary potentials in supergravity, is not so surprising. Nevertheless, in certain cases the lowering of the potential described here may help in extending the inflationary phase.

\begin{figure}[h!] 
\centering
\mbox{\subfigure{\includegraphics[width=0.49 \linewidth]{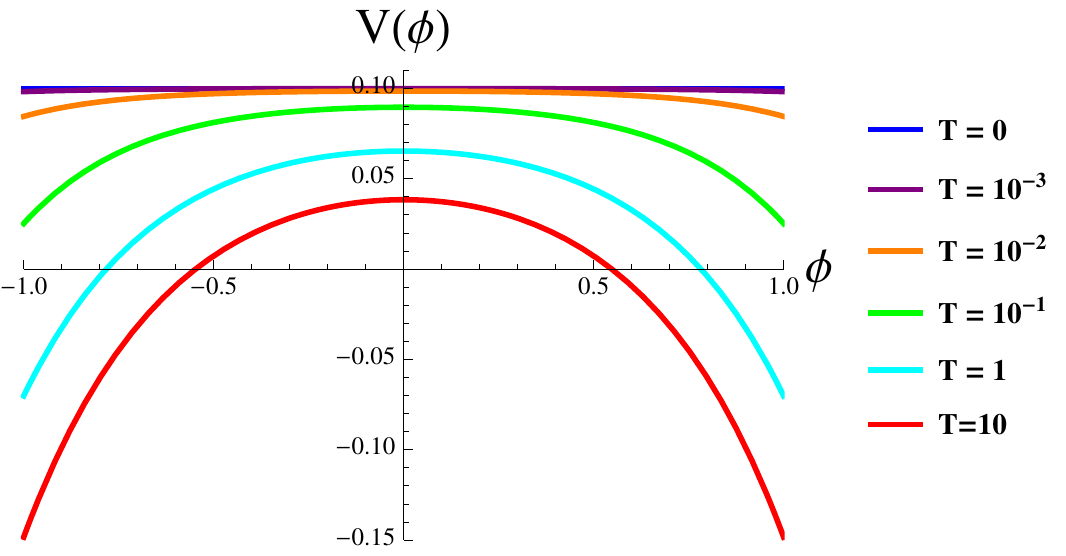}}\quad
\subfigure{\includegraphics[width=0.49 \linewidth]{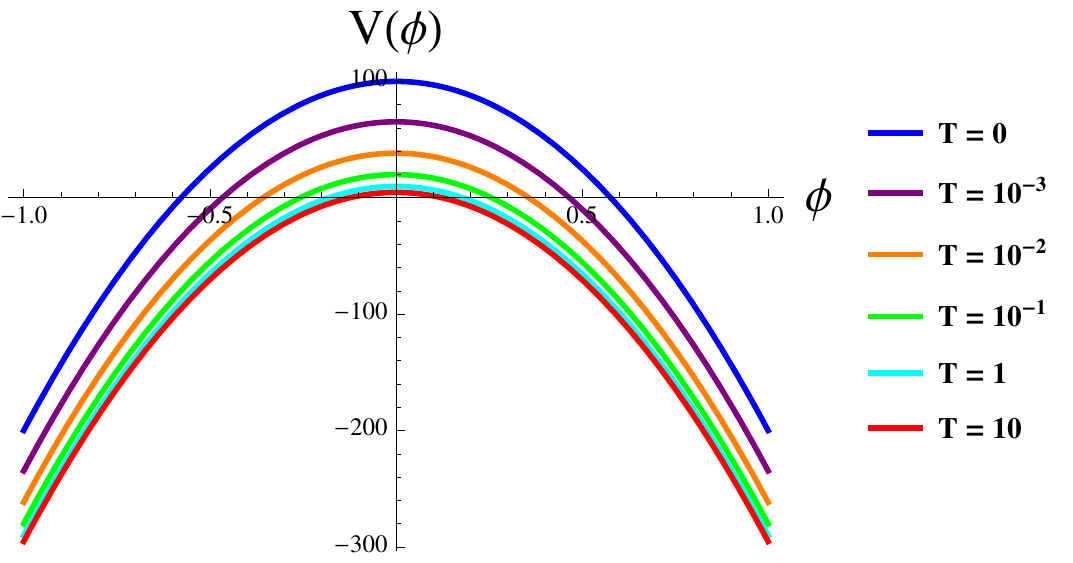}}}
\caption{Effect of increasing $T>0$ for an originally flat potential $-P_{Fl}(0, \phi)$ for $V_0=0.1$ (left) and for an unstable potential $-P_{Lin}(0, \phi)$ for $c=10$ (right).}\label{FigurePot1}
\end{figure}

\begin{figure}[h!] 
\centering
\mbox{\subfigure{\includegraphics[width=0.49 \linewidth]{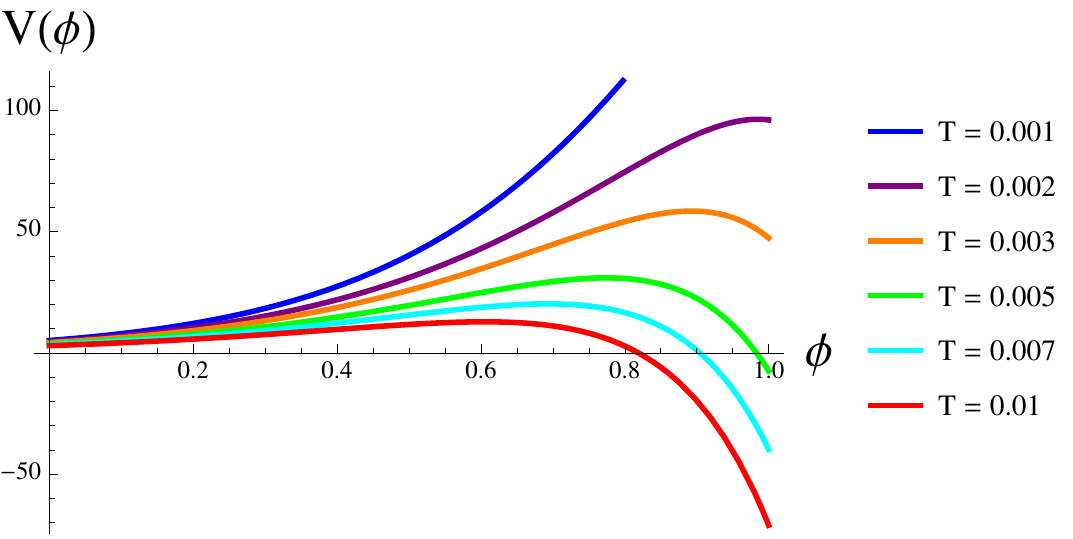}}\quad
\subfigure{\includegraphics[width=0.49 \linewidth]{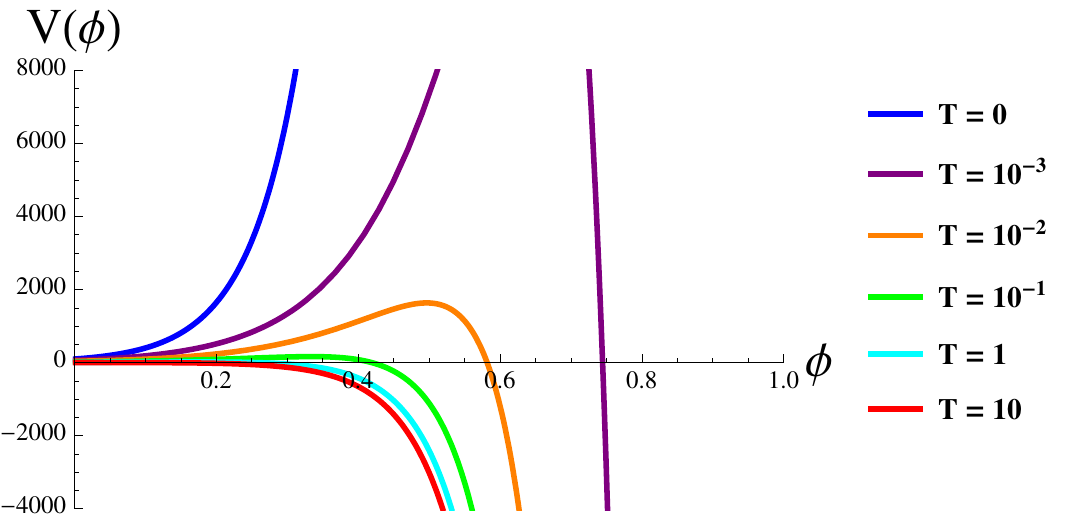}}}
\caption{Effect of increasing $T>0$ for the exponential potentials $-P_{CanK}(0, \phi)$ for $c=3$ (left) and on $-P_{QuadK}(0, \phi)$ for $c=10$ (right).} \label{FigurePot2}
\end{figure}


\section{Negative $T$}
\label{csblowup}

Up to now, we have been imposing $T>0,$ which corresponds to the standard expected sign for the first non-trivial correction $X^2$ to the standard kinetic term $X$ \cite{Adams:2006sv}. This had the consequence of reducing the number of possible solutions for the auxiliary field to only one, by virtue of the consistency condition \eqref{Freal}. However, in order to understand the structure of the theory better, it is interesting to also consider the $T<0$ case, where, as we will see, three branches of the theory can co-exist.

\label{Tneg}
\subsection{Three branches}
From Table \ref{examplestable} above, it is easy to see that the cubic for $F$ in each of the examples considered here is of the form
(\ref{cubic_general}) (for $\xi = 0$)
\begin{eqnarray}
0 & = & F^3 + p F + q,\\
0 & = &F^3 + \frac{C_1(\phi) F}{32 T} (1 + 32 XT) + \frac{C_2(\phi)}{32 T}.
\end{eqnarray}
Assuming as before that $c, V, D_A W, K_{A A^\star} > 0$, so that $C_1(\phi)$ and $C_2(\phi)$ are non-negative, allows us to examine the solutions to the cubic equation for the auxiliary field $F$ in detail. There are three solutions, given explicitly below, and they are not always all real. Which solution is real depends on the relative values of $C_1(\phi), C_2(\phi), X$ and $T$. 

Defining the discriminant $D = \frac{q^2}{4} + \frac{p^3}{27}$ as above, the three regions to consider are:
\begin{itemize}
\item {\bf Region 1}: $p<0; D<0$: This region is defined by the interval $0 <X < - \frac{1}{32 T} - \left ( \frac{27}{4} \frac{C_2(\phi)^2 }{C_1(\phi)^3 32^2T^2} \right)^{\frac{1}{3}}$ for fixed $T<0$. 
\item {\bf Region 2}: $p<0; D>0$: This region is given by $X >-  \frac{1}{32 T} - \left ( \frac{27}{4} \frac{C_2(\phi)^2 }{C_1(\phi)^3 32^2 T^2} \right)^{\frac{1}{3}}$ and $X <-  \frac{1}{32 T}$ for fixed $T<0$. 
\item {\bf Region 3}: $p>0; D>0$: This region is given by $X > -  \frac{1}{ 32 T}$ for fixed $T<0$. 
\end{itemize}
 
\noindent These regions are clearly seen in the plots below, for $C_1(\phi) = 1, C_2(\phi)= 0.1$.  The figure corresponds to the Linear case, but, keeping in mind that $F$ obeys an algebraic and not a differential equation, it is clear that the structure will be the same in general. We plot the three solutions for $F$, $F_B, F_M$ and $F_G$, colored in blue, magenta and green respectively. All three are real in Region 1 (where the labels should be read as $F_{B1}, F_{M1}$ and $F_{G1}$), but only $F_M$ is real in Region 2, and only $F_G$ is real in Region 3.
  
\begin{figure}[h!]
\vskip -3mm
\centering
\includegraphics[scale=0.8]{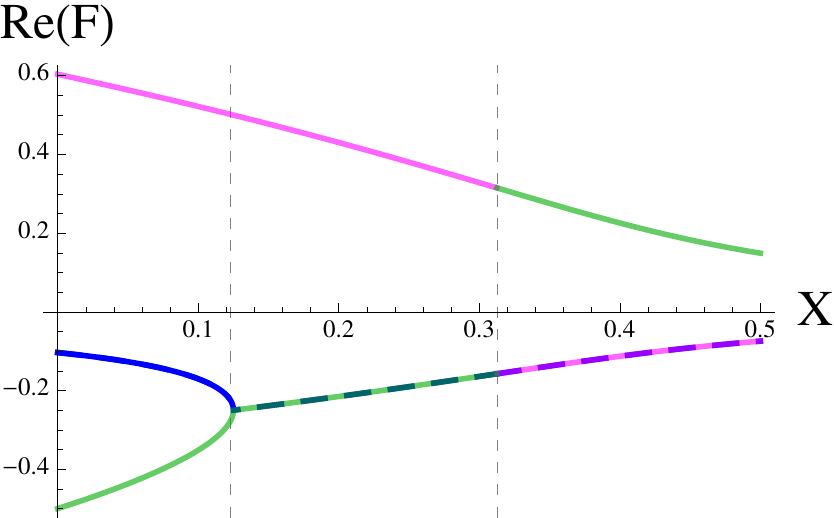}
\includegraphics[scale=0.8]{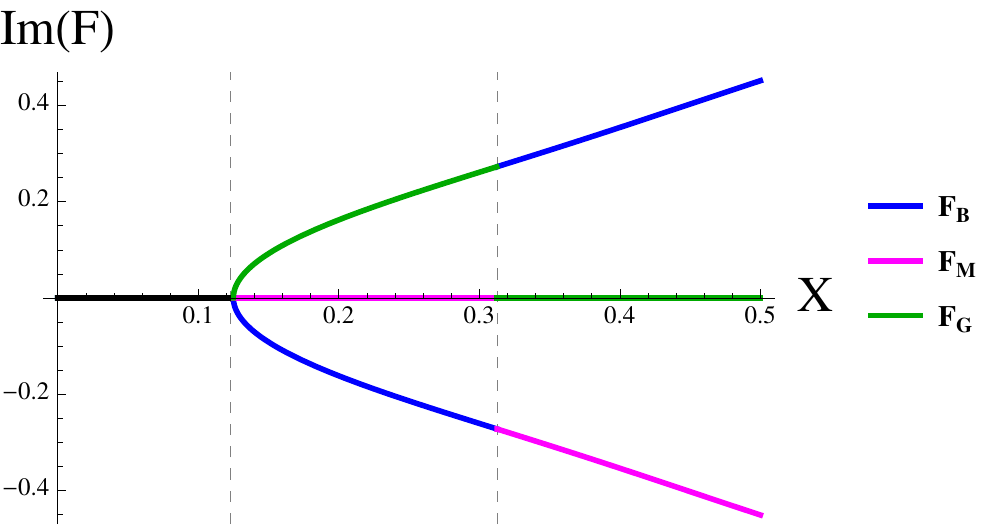}
\caption{The solutions to the cubic for $F$ as a function of $X$ for $\phi = 0.1, T = -0.1$. Note that for small $X$ all three solutions are real, and their imaginary parts thus overlap at $0$. }
\label{FsolnswrtX}
\end{figure}
 \begin{figure}[h!]
\vskip -3mm
\centering
\includegraphics[scale=0.8]{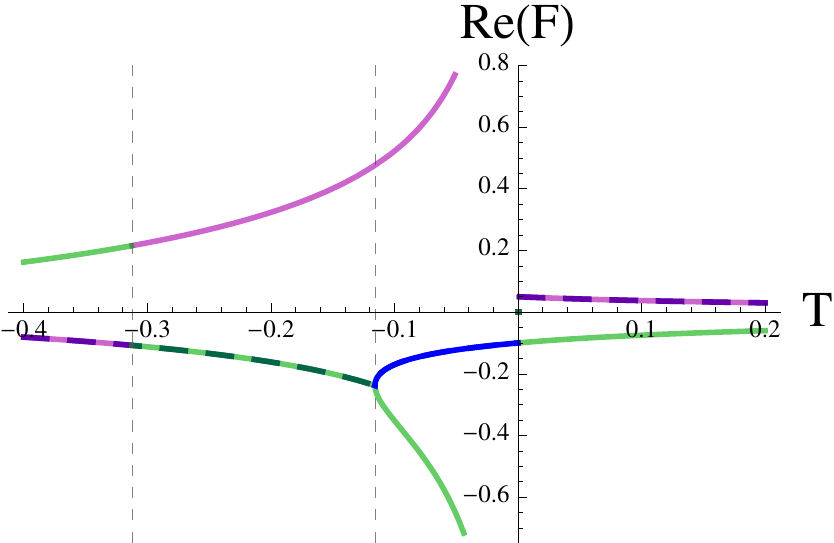}
\includegraphics[scale=0.8]{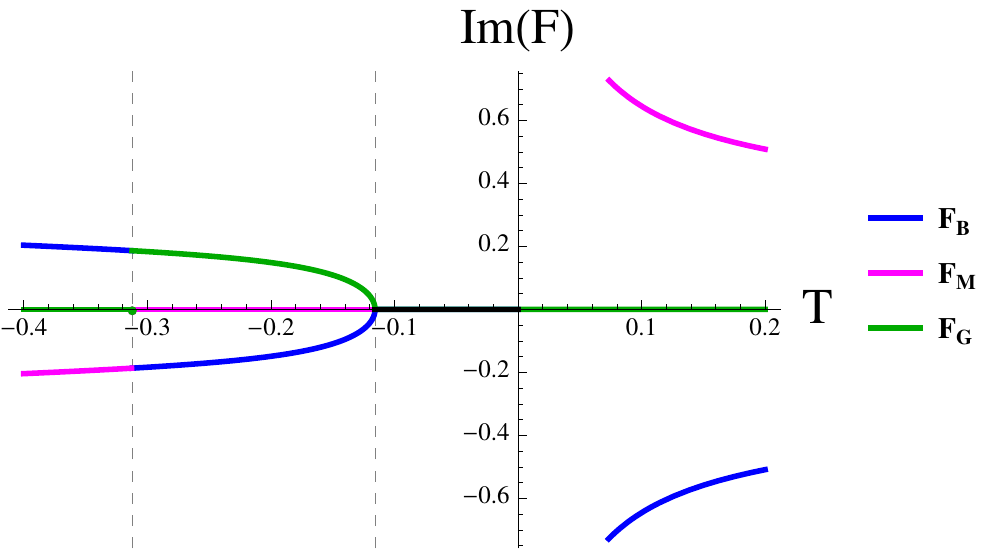}
\caption{The solutions to the cubic for $F$ as a function of $T$ for $\phi = 0.1, X = 0.1$.}
\label{FsolnswrtT}
\end{figure}

As expected, either one or all of the solutions are real in a given region. Here we have one real solution in regions 2 and 3, denoted by $F_M$ and $F_G$ respectively, and three real solutions in region 1. $F_B$ is only real in Region 1.  Note that $q < 0 \,\, \forall \,\,\,T<0$.  The explicit expressions,  valid in the regions when the respective solutions are real, are \footnote{Note that the full solutions to the cubic, found in Mathematica and plotted in figures \ref{FsolnswrtX} and \ref{FsolnswrtT}, are more complicated and show branch cuts as Mathematica always chooses the principal root for a given $n$th root (of which there are several in the full expressions).}
\begin{eqnarray}
F_{B1}(X, T) & = & 2 \sqrt{\frac{-p}{3}} \cos \left [ \frac{1}{3} \arccos \left ( \frac{ -9q}{2 \sqrt{3} (-p)^{3/2}} \right ) - \frac{2 \pi}{3} \right]\\
F_{M1}(X, T) & = & 2 \sqrt{\frac{-p}{3}} \cos \left [ \frac{1}{3} \arccos \left ( \frac{ -9 q}{2 \sqrt{3} (-p)^{3/2}} \right ) \right]\\
F_{G1}(X, T) & = & 2 \sqrt{\frac{-p}{3}} \cos \left [ \frac{1}{3} \arccos \left ( \frac{-9 q}{2 \sqrt{3} (-p)^{3/2}} \right ) + \frac{2 \pi}{3}\right]\\
F_{M2}(X, T) & = & -\frac{p}{3} (-D^{1/2} - \frac{q}{2})^{-1/3} - (-D^{1/2} - \frac{q}{2})^{1/3}\\
F_{G3}(X, T) & = & \frac{p}{3} (D^{1/2} + \frac{q}{2})^{-1/3} - (D^{1/2} + \frac{q}{2})^{1/3}
\end{eqnarray}

We can also plot the structure of the solutions as a function of $T$, keeping $X$ fixed -- see Fig. \ref{FsolnswrtT}. From the figure, we can see that $F_G$ is the solution that is real when $T>0$ (for real $p$). 

We note a few features that the graphs above reveal: even though there are three real solutions for small $X$ in the case where $T<0,$ only one of them remains real as the field velocity $X$ increases. Also, seen as a function of $T$, that solution (the one that is real for all negative $T$, i.e. the top line on the far left in Fig. \ref{FsolnswrtT}) does not match onto the solution that is real when $T$ becomes positive. Another interesting feature is the branching of the blue and green solutions occurring at negative $T.$ This makes one wonder whether, evolving from the ordinary (green) branch at small positive $T,$ one can actually keep evolving past the branching point at $T<0.$ We will now show that this is impossible, as a singularity is encountered before the branching point is reached.

\subsection{A singularity in the speed of sound}

The preceding discussion provides motivation to study the behavior of the $F_{B1}$ branch (which is smoothly connected to the ordinary branch at $T>0$) in more detail in the negative $T$ region. The fact that this solution for $F$ becomes imaginary when the field velocity increases sufficiently suggests that there may be some pathology associated with this branch in the negative $T$ region. We will now show that this is indeed the case, and that there occurs a blowup in the square of the speed of sound $c_s^2.$ 
\begin{figure}[h!]
\vskip -3mm
\centering
\label{cs2plot}
\includegraphics[scale=0.75]{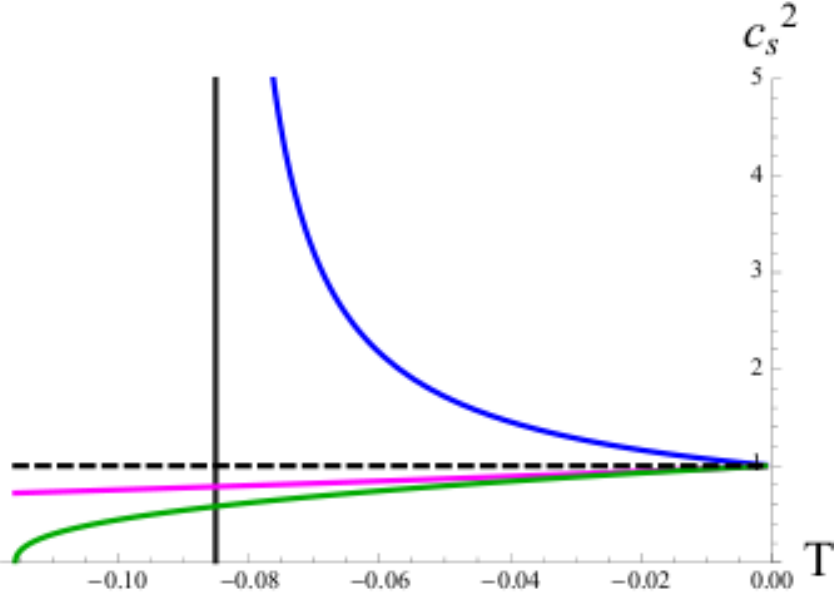}
\includegraphics[scale=0.75]{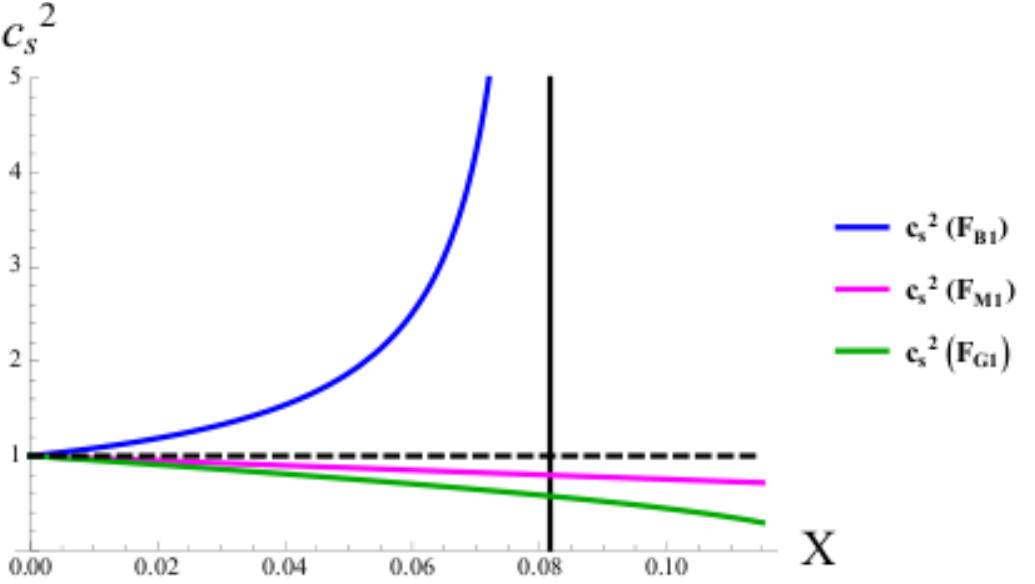}
\caption{$c_s^2$ for $c=0.1, \phi = 0.1$ as a function of $T$ (left), with $X = 0.1$, and as a function of $X$ (right), with $T = -0.1$. Note that the range of Region 1 is different for $X$ and $T$.}
\end{figure}
In fact, we can show that this blowup always takes place in Region 1: 
The general form of the Lagrangian and cubic is
\bea
\nonumber P(Z) & = & Z + 16 T Z^2  + 2 C_2 C_1^{-2} F + C_1^{-1} F^2 (1+ 32T Z)+ 16 T C_1^{-2} F^4    + 3 e^K |W|^2,\\
0 & = & F^3 + \frac{C_1 F }{32 T} (1 + 32T Z ) + \frac{C_2}{32 T} \\
C_1 &=& e^{-K/3} \frac{K^{,AA^\star}(D_A W)^\star}{D_A W} \\ C_2 &=& \frac{[K^{,AA^\star}(D_A W)^\star]^2}{D_A W}
\eea
where $Z = -K_{AA^\star} \partial A \cdot \partial A^\star$. Then
\bea
P_Z & = & 1 + 32T Z   + 32T C_1^{-1} F^2,
\eea
where we have used the cubic for $F$ to cancel the $F_Z$ terms. 
Using these we can evaluate $\rho$ and $\rho_Z$:
\bea
\rho & = & 2Z P_Z - P\\
\nonumber  & = & Z +48T Z^2 + 32T Z C_1^{-1} F^2 - C_1^{-1} F^2  - 2 F C_1^{-2}(C_2 + 8 T F^3 ) - 3 e^K |W|^2\\
\rho_Z & = & 1 + 96 T Z  + 32T C_1^{-1} F^2 + 128T C_1^{-1} F F_Z Z.
\eea
We can rewrite the $F_Z$ term by using the derivative of the cubic to get
\be
F_Z  = - \frac{32T C_1 F }{(C_1 +32T C_1 Z + 96 T F^2 )}
\ee
and write
\be
\rho_Z = 1 + 96T Z +  32T C_1^{-1} F^2 - \frac{(64T)^2 F^2 Z}{(C_1 + 32T C_1 Z + 96T F^2 )}.
\ee
Zeroes of $\rho_Z$ correspond to blowups in the squared speed of sound $c_s^2 = \frac{P_Z}{\rho_Z}$. In order to find the zeroes, we would like to solve the equation
\be
(64T)^2 C_1^{-1} F^2 Z  =  (1 + 96T Z  + 32T C_1^{-1} F^2 )(1 + 32T Z  + 96T  C_1^{-1} F^2 )
\ee
Let $Q = 1 + 32T Z  + 32T C_1^{-1} F^2$. Then this equation becomes
\be
0  =  Q (Q + 64T  Z  + 64T  C_1^{-1} F^2). 
\ee
Solutions are given by $Q = 0$ and $Q = - 64T Z  - 64T C_1^{-1} F^2$. Both are only possible when $T  < 0$. $Q = 0 \rightarrow C_2(\phi) = 0$ upon using the cubic, so we discard it. The other solution is
\bea
96  T ( \frac{F^2}{C_1(\phi)} + Z) & = & - 1\\
\Rightarrow F & = & - \frac{3 C_2(\phi)}{2 C_1 (\phi) } = -\frac{3}{2} K^{,AA^\star} (D_A W)^\star;\\
\label{X0} \Rightarrow Z & = & - \frac{1}{96  T} - \frac{9 C_2^2}{4 C_1^3}. 
\eea 
Let $Z = X$ i.e. $Y = 0$. Now consider that Region 1 (where there are three real roots of the cubic) is bounded by $X = 0$ and $D=0$ where 
\bea
D & =& \left (\frac{C_2}{64 T}\right )^2 + \left (\frac{C_1(1 + 32T X )}{96T} \right)^3.
\eea
At the point where $\rho_X = 0$ (at that point we write $X=X_0$, given by (\ref{X0})), 
\be
D = - \frac{3}{4} C_2^2 X_0^2  - \frac{8}{27} C_1^3 X_0^3 \,\, < \,\, 0,
\ee
so we find that the blowup always occurs in Region 1. There remains the question of which branch it occurs on. This is answered by noting first that $F_0 <0$, so it cannot be on the $F_{M1}$ branch. Secondly, evaluating $F_X$ at the blow-up point one finds $F_X = \frac{F}{2X_0} <0$, which implies that it must be on the $F_{B1}$ branch (see Figure \ref{FsolnswrtX}). This is exactly the branch that meets the $T>0$ solution at $T=0$, as we found from the plots above, Figure \ref{cs2plot}. 

Superluminality of the propagation of fluctuations was pointed out in \cite{Adams:2006sv} as resulting from the presence of higher-derivative corrections with negative sign. The sign of these terms is given here by the sign of $T$, so that one might expect such problems, but it is interesting to see that $c_s^2$ is immediately greater than 1, and that the blow-up to infinity happens so soon when $T$ becomes negative. We expect the theory in this regime to have problems with causality and locality. Because this theory is an attempt at a full supergravity embedding of theories with higher-power kinetic terms, and has no inherent constraint on $T$, we expect that there is an additional constraint at the string theory level which should be inherited by the low-energy theory. 

\begin{figure}[h!]
\vskip -3mm
\centering
\label{cs2regions23}
\includegraphics[scale=1]{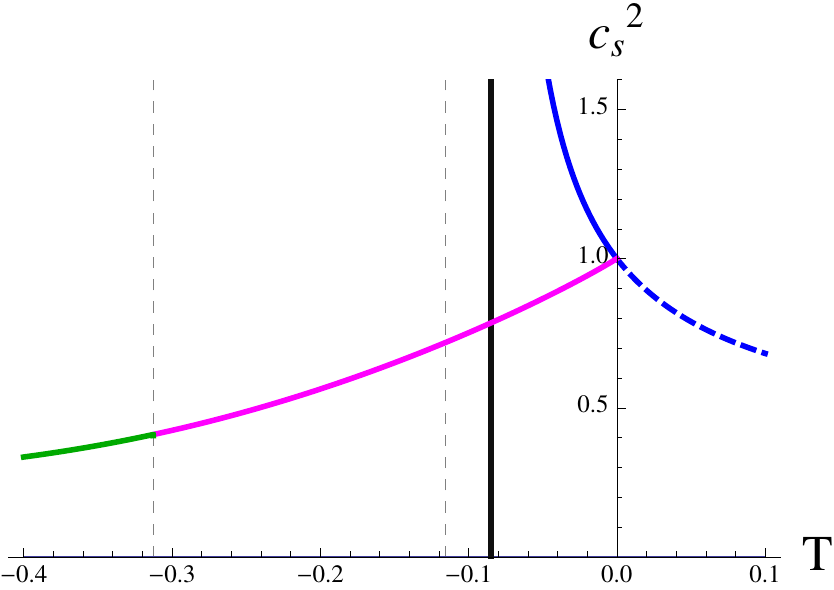}
\raisebox{2cm}{\includegraphics[scale=1]{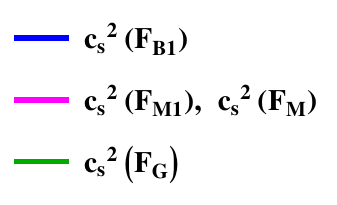}}
\caption{$c_s^2$ for $c=0.1, \phi, X = 0.1$ as a function of $T$, showing the discontinuity at $T=0$ for the nonsingular branch.} 
\end{figure}

On the other hand if one begins on the $F_M$ or $F_G$ solution in Region 1, one can approach $c_s^2 = 1$ without problems as $T \rightarrow 0^{-}$. The real branch for $c_s^2$ is given by $F_G$ in Region 3 and $F_M$ in Region 2, and this is smoothly connected to the $F_M$ solution in Region 1. However, it is clear from the plot that this solution is not smoothly connected to the $T \geq 0$ solution, so it is not possible to move smoothly between these solutions by varying $T$. This situation is shown in Figure \ref{cs2regions23}.

\section{Discussion}

In this work we have studied the effective field theory approach to cosmology in ${\cal N}=1$ 4-dimensional supergravity, focussing on the effects of the leading higher-derivative kinetic term corrections. We would like to emphasise that from a string-theoretic perspective such terms are expected to be present on general grounds. Moreover, the cosmic microwave background data from the PLANCK satellite still leaves considerable room for the presence of such terms, as the experimental error on the associated equilateral and orthogonal non-Gaussianity parameters $f_{NL}^{equil}=-42 \pm 75$ and $f_{NL}^{ortho}=-25 \pm 39$ (at $2\sigma$) \cite{Ade:2013uln} are still quite large. 

In supergravity, the leading and simplest higher-derivative terms are of the form \cite{Koehn:2012ar}
\be
T (\partial A)^2 (\partial A^\star)^2-2T e^{K/3} \partial A \cdot \partial A^\star FF^\star + T e^{2K/3} (FF^\star)^2.
\ee
The role of the auxiliary field $F$ is crucial in this context, as it leads to corrections to both the ordinary kinetic term and, most importantly, to the potential. Even in the regime where the coefficient of the square of the ordinary kinetic term is small, there can be dramatic changes occurring in the potential, which drastically modify any possible inflationary dynamics. We have restricted our analysis to the case where $T$ is a constant. The general formalism \cite{Koehn:2012ar} would allow $T$ to also contain derivatives of the fields, e.g. $T$ could contain a factor $(-\partial A \cdot \partial A^\star)^n \supset X^n$ with $n>0.$ As is evident from the expression above, such a term would then also yield corrections to the $X^{n+1}$ and $X^n$ terms -- however, it would not modify the form of the potential term (except of course for the change it induces in the solution for $F$). In an inflationary context, the non-derivative $T$ corrections are thus the most important ones. 

For $T \geq 0$, we gave the general solution for $F$. We showed that the corrections can be large even when $T$ is small -- in particular, the corrections cannot be ignored when 
\begin{eqnarray}
32 T e^K K^{A A^\star} |D_A W|^2 &\gtrsim&1.
\end{eqnarray}
We then compared the $T$-dependent kinetic and potential term contributions, finding  that the T-dependent contribution to the potential will be more significant than the T-dependent contribution to the kinetic part of the action. We have also verified that the attractor formalism developed in \cite{Franche:2009gk} remains valid for these supergravity $P(X , \phi)$ actions. We considered several explicit examples in detail, confirming that for these even very small increases in $T$ can result in significant changes to the potential and to the inflationary dynamics.

Our findings suggest that it is essential to take into account the full supergravity description of higher-derivative terms, rather than assuming that a small parameter expansion is possible in which the only relevant terms are $X^n$ etc. If a potential is already unstable, the higher-derivative $T$-dependent corrections tend to make it even more unstable. In the inflationary context, the higher-derivative terms play a more welcome role when the potential curves upwards too steeply (i.e. when $V_{,\phi\phi}>0$), since in such a situation they can lower and flatten the potential. In addition the higher-derivative kinetic term slows the field down, as expected in non-canonical inflationary scenarios. On the other hand the effect of higher-derivative corrections on the potential may be useful in ekpyrotic models also, where a steep negative potential is required \cite{Khoury:2001wf, Lehners:2008vx} -- we leave such an investigation for future work.

For $T <0$, we gave general expressions for the three solutions to the cubic for $F$ in section \ref{Tneg}, and determined which is real for the three relevant $T <0$ regions defined by the relative values of $T$ and $X$. However,  the system seems to be blocked from the region with three real solutions for $T <0$ by a blow-up of $c_s^2$, the speed of propagation of perturbations, which should have $c_s=1$ as an upper limit. 

This study represents an important step forward in the understanding of how higher-order kinetic terms should be embedded in a consistent (cosmological) supergravity context. It will be interesting to look for set-ups in which these effects lead to a useful model of inflationary or ekpyrotic cosmology, consistent with the available data but not suffering from excessive fine tuning. We hope to be able to return to this question in the future.

\section*{Acknowledgments}

We would like to thank Lorenzo Battarra, Michael Koehn and Burt Ovrut for useful discussions. The authors gratefully acknowledge the support of the European Research Council via the Starting Grant Nr. 256994 ``StringCosmOS''.

\appendix

\section{Stability of our examples}

In the present paper we are interested in solutions to the equations of motion implied by the Lagrangians (\ref{PCanK})-(\ref{PLin}) in the case $\xi = 0$. For this to be a sensible undertaking, we must analyze whether or not the systems are stable around $\xi = 0$. Below, we find the conditions under which we have stability, but find that for the ``Flat'' case, perhaps unsurprisingly, the potential for the second scalar $\xi$ is always unstable. 

When $T \neq 0$, $F$ can be complex and it is not immediately clear which solution to the cubic is correct. With the exception of section \ref{csblowup} where we consider all solutions, we will always choose the {\it ordinary branch}, which is the solution that connects smoothly to the $T=0$ solution \cite{Koehn:2012ar}. We will analyze our four examples in turn.

\begin{figure}[htp]
\centering
\mbox{\subfigure{\includegraphics[width=0.49 \linewidth]{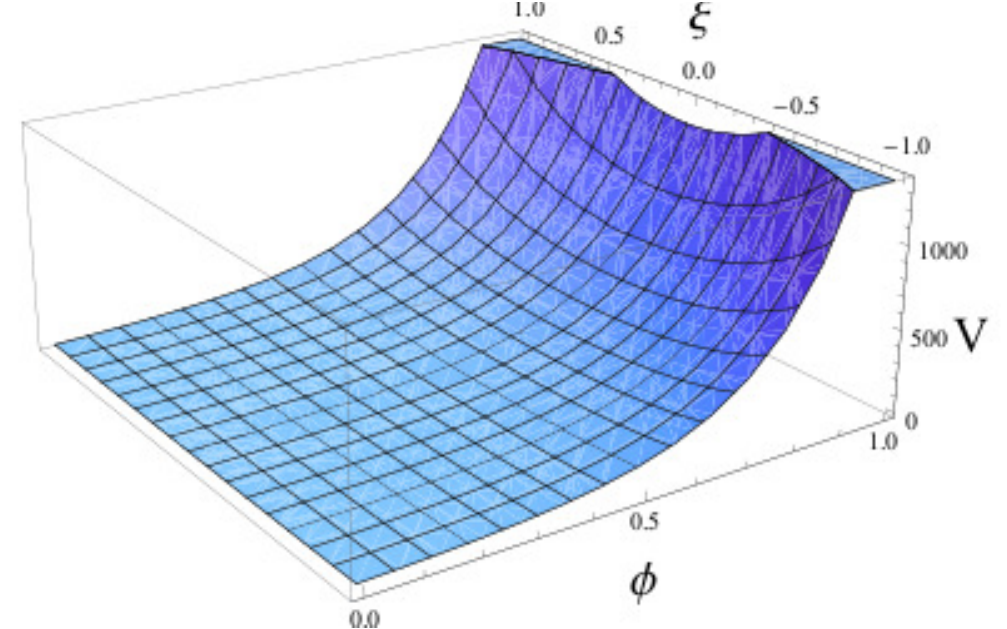}}\quad
\subfigure{\includegraphics[width=0.49 \linewidth]{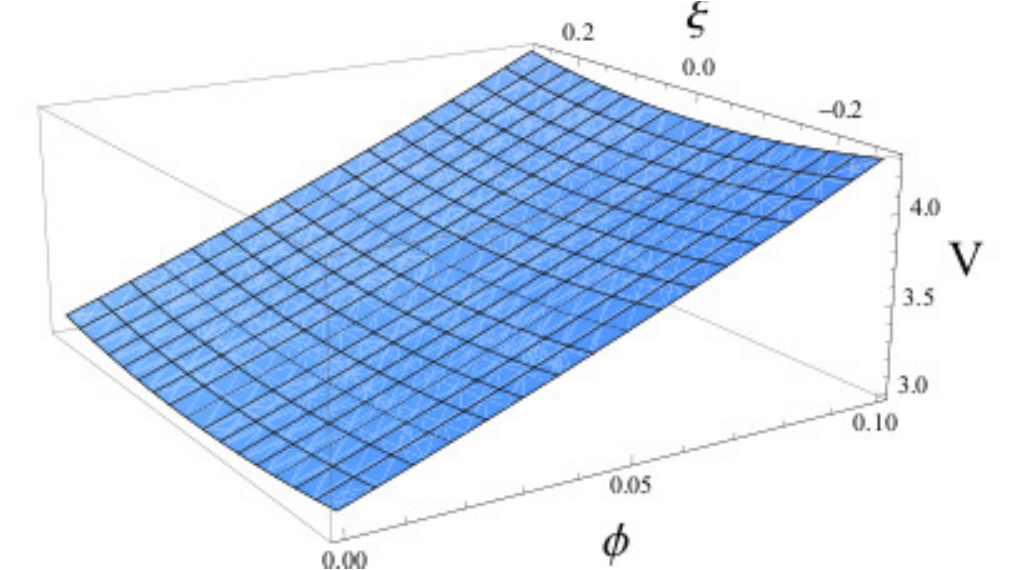}}}
\caption{$\xi$ potential from $P_{CanK}$ for $T=0$ (left) and $T = 0.01$ (right), with $c = 3$.}
\label{CanKStabPlots}
\end{figure}

The Lagrangian for the Canonical K\"ahler potential is given in (\ref{PCanK}). For  $T=0$, we find
\begin{eqnarray}
P_{CanK}(X, Y, \phi, \xi) & = & X + Y + e^{(\phi^2 + \xi^2)/2} e^{\sqrt{2} c \phi} \left [ 3 - \left (\left ( c + \frac{\phi}{\sqrt{2}}\right )^2 + \xi^2 \right) \right ]\\
\nonumber & \approx & X + Y + e^{\sqrt{2} c \phi + \frac{\phi^2}{2}} [ 3 - (c + \frac{\phi}{\sqrt{2}})^2] + \frac{\xi^2}{2} e^{\sqrt{2} c \phi + \frac{\phi^2}{2}} [2 - (c + \frac{\phi}{\sqrt{2}})^2] \\ \nonumber && + {\cal O} (\xi^3),
\end{eqnarray}
from which we see that the potential for $\xi$ is positive and stable as long as $(c + \frac{\phi}{\sqrt{2}})^2 > 2$. It is more complicated to evaluate the full potential for $\xi$ explicitly in the $T \neq 0$ case, because this involves substituting in a complicated expression for a generally complex $F$. However, we can deal with this case both by expanding $T,$ and numerically. Expanding $F = F_0 + F_1 T,$ where $F_0 = - e^{K/3} K^{A A^\star} (D_A W)^\star$ etc as in (\ref{smallTapprox}), we find that for $T\ll 1$ the potential is given by 
\begin{eqnarray}
V_{CanK}(\phi, \xi) & \approx & \frac{1}{2} e^{\sqrt{2} c \phi} e^{\phi^2/2} (-6 + 2 (c + \frac{\phi}{\sqrt{2}})^2) - 16 (e^{\sqrt{2} c \phi} e^{\phi^2/2})^2 (c + \frac{\phi}{\sqrt{2}})^4 T
 \\ \nonumber && + \left [ \frac{1}{2} e^{\sqrt{2} c \phi} e^{\phi^2/2} (-2 + (c + \frac{\phi}{\sqrt{2}})^2) - 16 e^{2 \sqrt{2} c \phi} e^{\phi^2}(c + \frac{\phi}{\sqrt{2}})^2 ((c + \frac{\phi}{\sqrt{2}})^2 + 1) T\right ] \xi^2.
 \\ \nonumber && + {\cal O} (T^2,\xi^3)
\end{eqnarray}
From this approximation, a good one at very small $T$, we see that the $T$-dependent contributions to the potential have negative sign and will therefore make the potential for $\xi$ flatter and eventually unstable. We see this numerically as well; the potential is extremely flat in the $\xi$ direction compared to the $\phi$ direction - see Figure \ref{CanKStabPlots}. As $T$ increases, the $\phi$ value at which the potential flips becomes smaller and smaller. 

\begin{figure}[htp]
\centering
\mbox{\subfigure{\includegraphics[width=0.49 \linewidth]{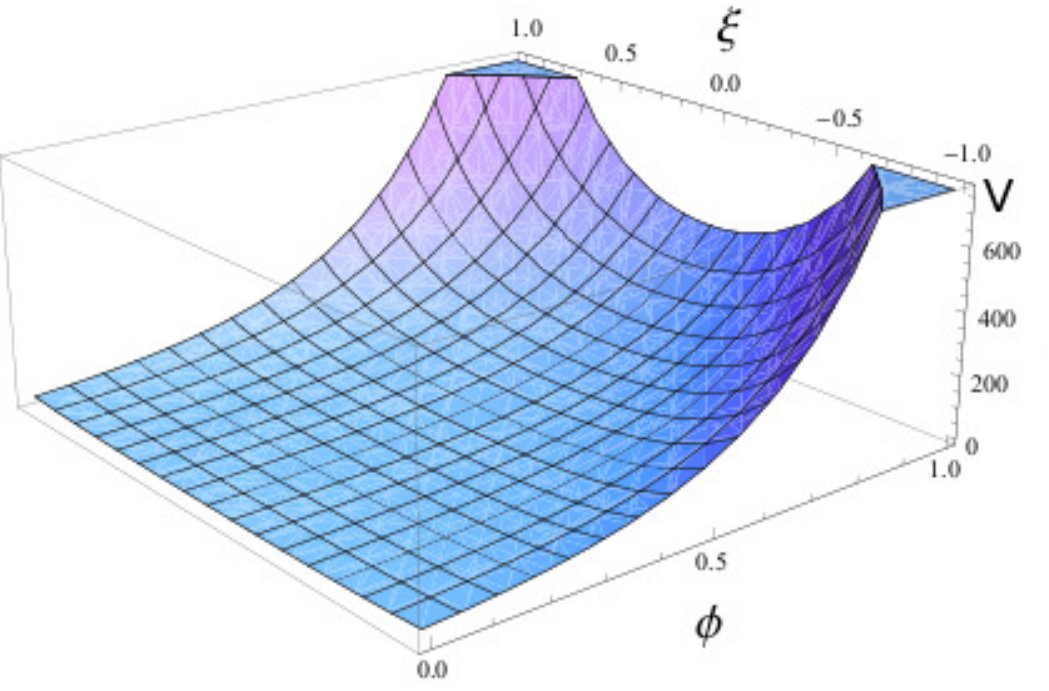}}\quad
\subfigure{\includegraphics[width=0.49 \linewidth]{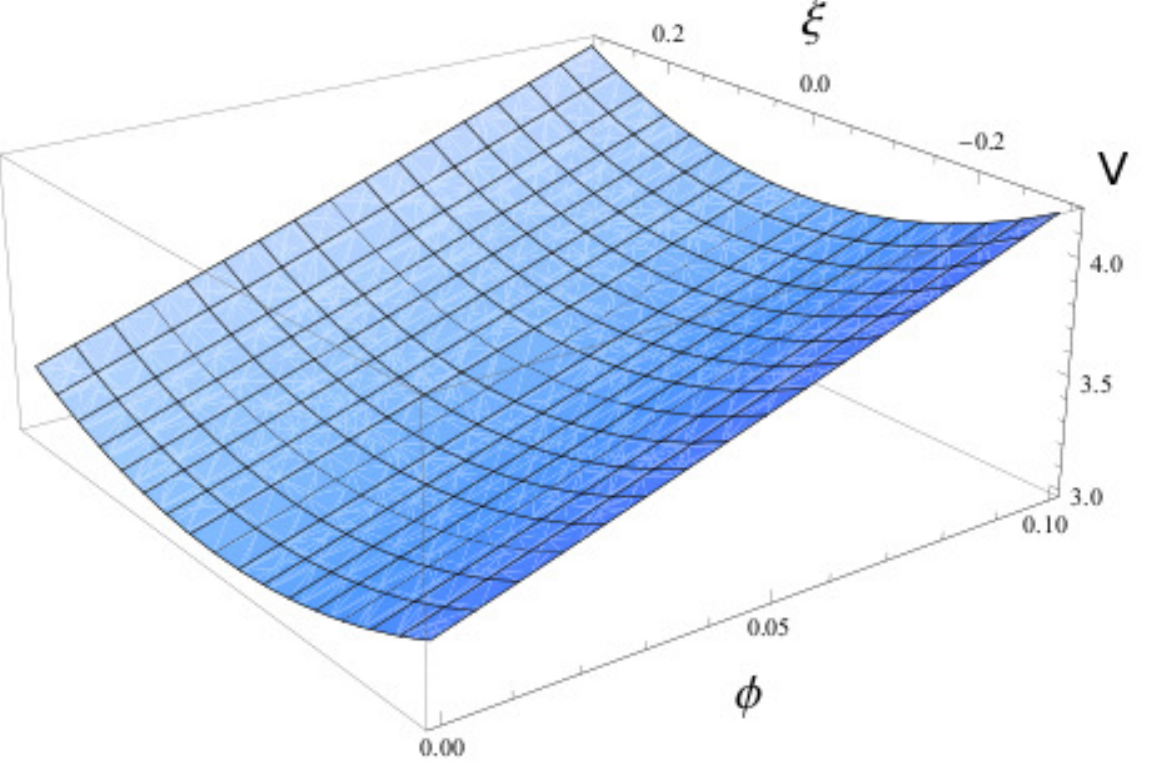}}}
\caption{$\xi$ potential from $P_{QuadK}$ for $T=0$ (left) and $T= 0.01$ (right), with $c = 3$.}
\label{SteepStabPlots}
\end{figure}

The Lagrangian for the Quadratic K\"ahler potential is given in (\ref{PQuadK}). For  $T=0$, we find
\begin{eqnarray}
V_{QuadK}(\phi, \xi)_{T=0} & \approx &  e^{\sqrt{2} c \phi}(c^2- 3) + \xi^2 e^{\sqrt{2} c \phi} (c^2 -1) + {\cal O}(\xi^3)
\end{eqnarray}
from which we see again that the potential for $\xi$ is positive as long as $c$ is large enough. For $ T\neq 0$, and working to linear order in $T$ in our expansion for $F$ as above, we find 
\begin{eqnarray}
\nonumber V_{QuadK}(\phi, \xi) & \approx & e^{\sqrt{2} c \phi} \left [ c^2 - 3 - 16 c^4 e^{\sqrt{2} c \phi} T +  (c^2 - 1) \xi^2 - 32 e^{\sqrt{2} c \phi} c^2  (c^2  + 2) T \xi^2\right] \\  && + {\cal O} (T^2,\xi^3).
\end{eqnarray}
In this case we see that the $T$-dependent terms are all negative. For $c$ large enough to give a positive mass term for $\xi$, the potential will be dominated by the third term, $-16 c^4 e^{\sqrt{2} c \phi} T$.  This is seen in the numerical results as well, where the potential is flatter when $T$ is switched on; see Figure \ref{SteepStabPlots}. Note however that the potential is  deeper in the Quadratic K\"ahler potential case than the Canonical K\"ahler case, for the same values of $c$ and $T$.  A full numerical analysis shows that, for a given value of $c$, a large enough $T$ will flip the potential for $\xi$. For $c=3$ and $T = 0.01$, this flip occurs at $\phi \sim 0.6$.

\begin{figure}[htp]
\centering
\mbox{\subfigure{\includegraphics[width=0.49 \linewidth]{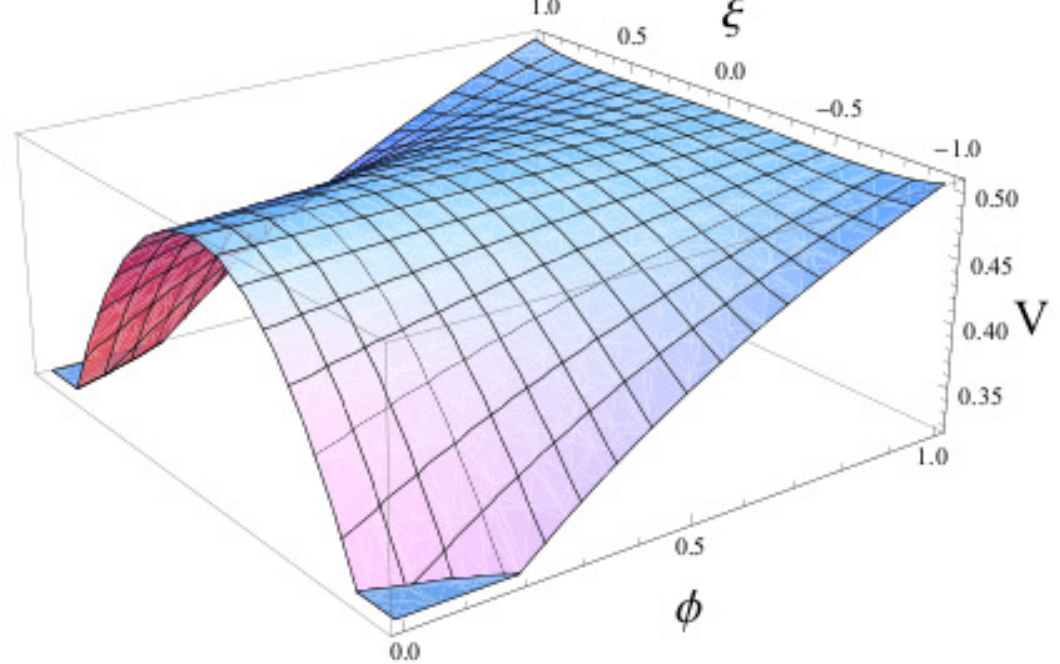}}\quad
\subfigure{\includegraphics[width=0.49 \linewidth]{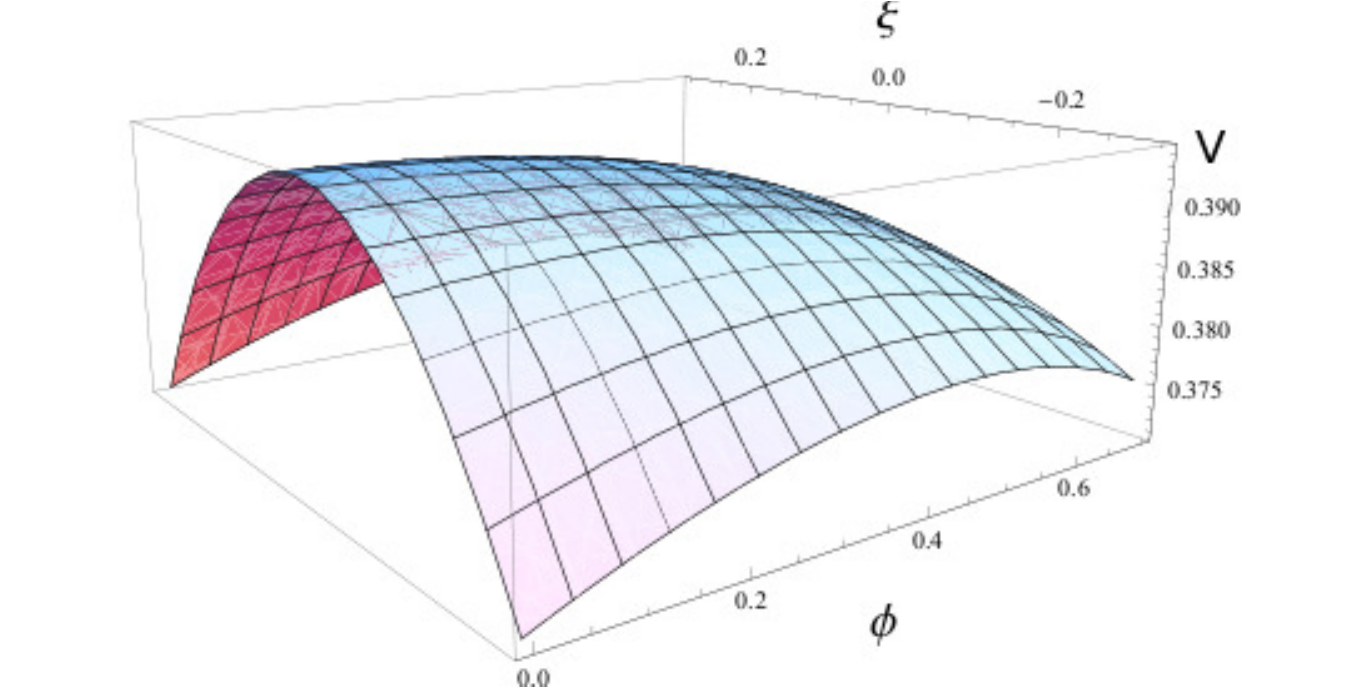}}}
\caption{$\xi$ potential from $P_{Fl}$ for $T=0$ (left) and $T=0.001$ (right), with $V_0 = 0.5$.}
\label{FlatStabPlots}
\end{figure}

\begin{figure}[htp]
\centering
\mbox{\subfigure{\includegraphics[width=0.49 \linewidth]{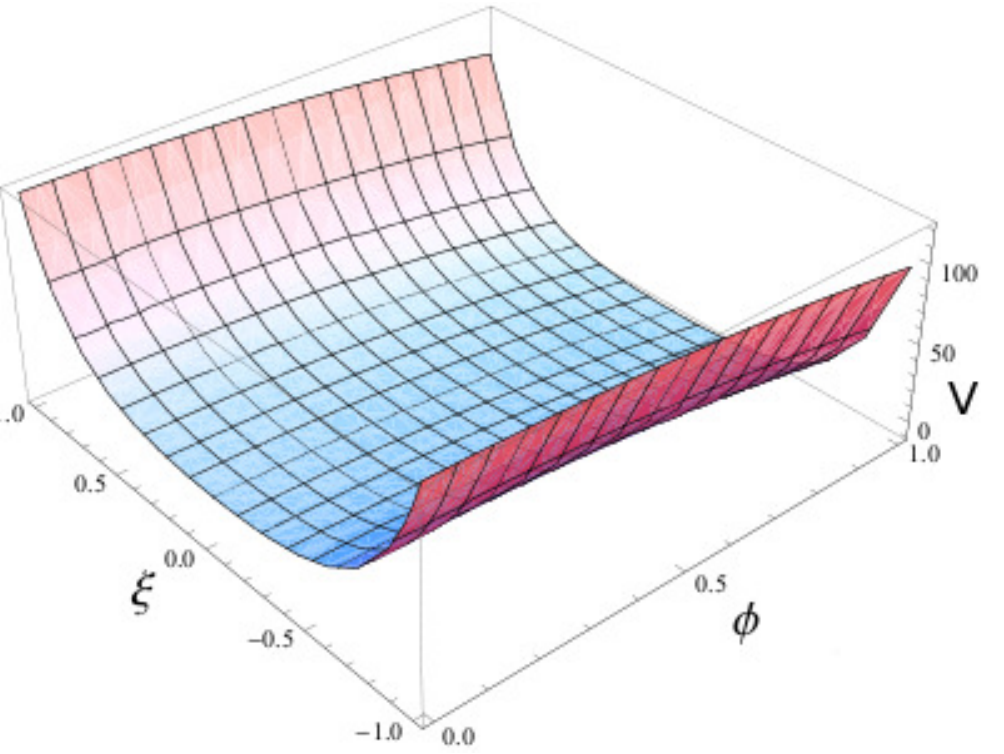}}\quad
\subfigure{\includegraphics[width=0.49 \linewidth]{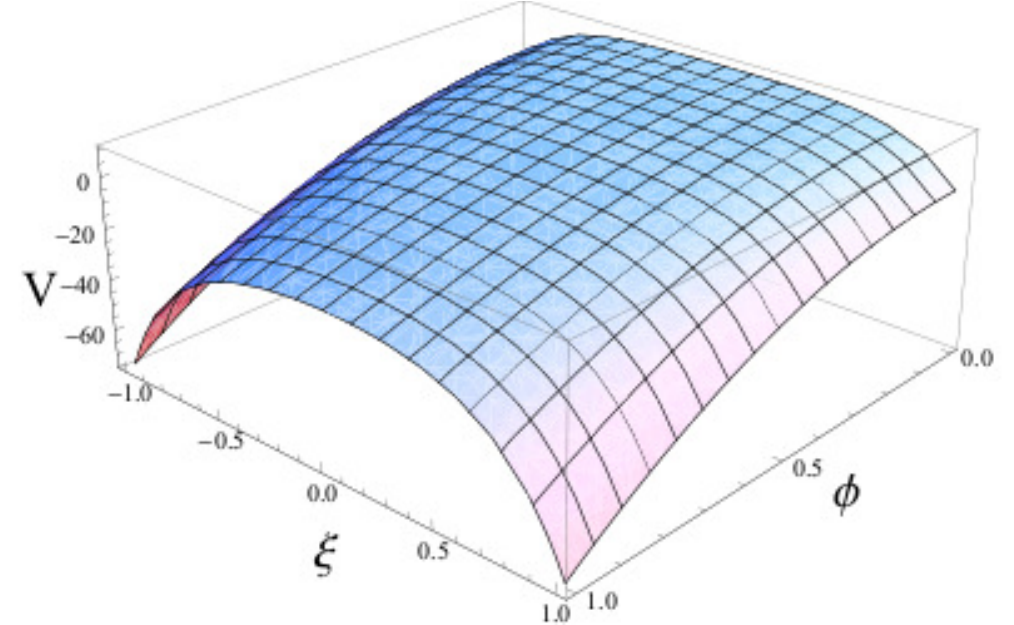}}}
\caption{$\xi$ potential from $P_{Lin}$ for $T=0$ (left) and $T= 0.01$ (right), with $c = 3$. Note that the potential becomes negative when $T$ is switched on, with $c$ left the same.}
\label{LinStabPlots}
\end{figure}

The Lagrangian for the Flat potential is given in (\ref{PFl}). For  $T=0$, we find
\begin{eqnarray}
V_{Fl} (X, Y, \phi, \xi)_{T=0}& \approx & V_0 + \frac{V_0}{24} e^{-(\sqrt{2} + \sqrt{6}) \phi} (1 - 14 e^{\sqrt{6} \phi} + e^{2 \sqrt{6} \phi} ) \xi^2 .
\end{eqnarray}
We see from this that the potential for $\xi$ is unstable around $\xi = 0$ (until $\phi \sim 2.63$), as is clear from Figure \ref{FlatStabPlots}. In the small $T$ approximation, we find
\begin{eqnarray}
V_{Fl}(\phi, \xi) & \approx & V_0 - V_0^2 e^{ - 2 \sqrt{6}\phi} (1 + e^{\sqrt{6} \phi})^4 T
+  \frac{V_0}{24} e^{\sqrt{2} \phi (1 - \sqrt{3})}
 (1 - 14 e^{\sqrt{6} \phi} + e^{2 \sqrt{6}{ \phi}})\xi^2 \\&&
\nonumber  - \frac{V_0^2 (1 + e^{\sqrt{6} \phi })^2}{3}e^{- 2 \sqrt{6} \phi - \sqrt{2} \phi} (7 - 3 \sqrt{3} + 10 e^{\sqrt{6} \phi}+ (7 + 3 \sqrt{3}) e^{2 \sqrt{6} \phi}) T  \xi^2.
\end{eqnarray}
Clearly the $T$-dependent terms here tend to destabilize the $\xi$-potential further, as is seen in Figure \ref{FlatStabPlots}. For large enough $\phi$ the negative terms can be suppressed, but at small $\phi$ and $\xi$ the potential is unstable and becomes more so as $T$ is switched on. This is perhaps to be expected as $P_{Fl}$ was chosen to give an exactly flat potential when $\xi = 0$.


The Lagrangian for the Linear potential is given in (\ref{PLin}). For  $T=0$, we find
\begin{eqnarray}
V_{Lin} (X, Y, \phi, \xi)_{T=0} &\approx & c^2 (2- 3 \phi^2 ) + c^2 \xi^2 (3-\phi^2 ),
\end{eqnarray}
from which we find that the potential for $\xi$ is stable around $\xi = 0$ for sufficiently small $\phi$. This result carries over to the $T \neq 0$ case only when $c$ is sufficiently small, as we see from the small $T$ approximation: 
\begin{eqnarray}
\nonumber V_{Lin}(\phi, \xi) & \approx & 2 c^2 (1 - \frac{3}{2} \phi^2 - 32 c^2 T) + c^2 (3 - \phi^2 - 128c^2 T (3 + \phi^2)) \xi^2. 
\end{eqnarray}
Again the $T$-dependent terms here tend to destabilize the $\xi$-potential, as is seen in  Figure \ref{LinStabPlots}. 

\bibliographystyle{JHEP.bst}
\bibliography{ncisugra}
\end{document}